%% file: PSRJ2051_v3.tex
\documentclass[a4paper,fleqn,usenatbib]{mnras}
\usepackage{amsmath}
\usepackage{txfonts}
\usepackage[T1]{fontenc}
\usepackage{ae,aecompl}
\usepackage{graphicx}
\usepackage{amssymb}

\newcommand{\PBDmax}{1.41 \times 10^{-11}}
\newcommand{\PBDmin}{-2.03 \times 10^{-11}}
\newcommand{\PbGW}{-7.61 \times 10^{-14}}
\newcommand{\Pbdoppler}{-4.06 \times 10^{-21}}

\newcommand{\XDmax}{2.29 \times 10^{-13}}
\newcommand{\XDmin}{-5.08 \times 10^{-13}}
\newcommand{\xGW}{-2.67 \times 10^{-19}}
\newcommand{\xPM}{4.99 \times 10^{-17}}
\newcommand{\xdea}{-4.41 \times 10^{-17}}
\newcommand{\xdoppler}{-4.06 \times 10^{-21}}

\newcommand{\xQ}{See \cite{lvt+11}}
\newcommand{\psr}{PSR J2051$-$0827}

\title[21-year timing of \psr{}]{21-year timing of the black-widow pulsar J2051$-$0827}

\author[G. Shaifullah et al.]{%
G.~Shaifullah,$^{1,2}$\thanks{E-mail: golam@mpifr-bonn.mpg.de (GS)}
J.~P.~W.~Verbiest,$^{1,2}$
P.~C.~C.~Freire,$^{2}$
T.~M.~Tauris,$^{2,3}$
N.~Wex,$^{2}$
\newauthor 
S.~Os{\l}owski,$^{1,2}$
B.~W.~Stappers,$^{7}$
C.~G.~Bassa,$^{4}$
R.~N.~Caballero,$^{2}$
D.J.~Champion,$^{2}$
\newauthor
I.~Cognard,$^{6,5}$
G.~Desvignes,$^{2}$ 
E.~Graikou,$^{2}$
L.~Guillemot,$^{6,5}$
G.~H.~Janssen,$^{4}$
\newauthor
A.~Jessner,$^{2}$
C.~Jordan,$^{7}$
R.~Karuppusamy,$^{2}$
M.~Kramer,$^{2}$
K.~Lazaridis,$^{2}$
P.~Lazarus,$^{2}$
\newauthor
A.~G.~Lyne,$^{7}$
J.~W.~McKee,$^{7}$
D.~Perrodin,$^{8}$
A.~Possenti,$^{8}$
C.~Tiburzi$^{2,1}$
\\
$^{1}$Fakult\"{a}t f\"{u}r Physik, Universit\"{a}t Bielefeld, Postfach 100131, 33501 Bielefeld, Germany\\
$^{2}$Max-Planck-Institut f\"{u}r Radioastronomie, Auf dem H\"{u}gel 69, 53121 Bonn, Germany \\
$^{3}$Argelander-Institut f\"{u}r Astronomie, Universit\"{a}t Bonn, Auf dem H\"{u}gel 71, 53121 Bonn, Germany\\
$^{4}$ASTRON, The Netherlands Institute for Radio Astronomy, Postbus 2, {NL-7900} AA, Dwingeloo, The Netherlands\\
$^{5}$Station de radioastronomie de Nan\c{c}ay, Observatoire de Paris, CNRS/INSU, Universit\'{e} d'Orl\'{e}ans, 18330 Nan\c{c}ay, France\\ 
$^{6}$Laboratoire de Physique et Chimie de l'Environnement, CNRS, 3A Avenue de la Recherche Scientifique, 45071 Orl\'{e}ans Cedex 2, France\\
$^{7}$Jodrell Bank Centre for Astrophysics, School of Physics and Astronomy, The University of Manchester, Manchester M13 9PL, UK\\
$^{8}$INAF - Osservatorio Astronomico di Cagliari, Via della Scienza 5, 09047 Selargius (CA), Italy 
}

\date{Accepted XXX. Received YYY; in original form ZZZ}

\pubyear{2015}

\begin{document}
\label{firstpage}
\pagerange{\pageref{firstpage}--\pageref{lastpage}}
\maketitle

\begin{abstract}
Timing results for the black-widow pulsar J2051$-$0827 are presented, using a 21-year dataset from four European Pulsar Timing Array telescopes and the Parkes radio telescope. This dataset, which is the longest published to date for a black-widow system, allows for an improved analysis that addresses previously unknown biases. While secular variations, as identified in previous analyses, are recovered, short-term variations are detected for the first time. Concurrently, a significant decrease of $\sim2.5 \times 10^{-3}\ \rm{cm^{-3}pc}$ in the dispersion measure associated with \psr{} is measured for the first time and improvements are also made to estimates of the proper motion. Finally, \psr{} is shown to have entered a relatively stable state suggesting the possibility of its eventual inclusion in pulsar timing arrays.
\end{abstract}




\section{Introduction}\label{sec:intro}
Of the $\sim$2600 pulsars known today, roughly 10\% appear to have rotation-periods of the order of a few milliseconds and are known as millisecond pulsars (MSPs). Within the MSP population there exist a variety of configurations, however, most MSPs are found in binary systems. Among these, about 10\% are in tight, eclipsing binaries. Such systems are further classified into the black-widow systems, with very light companions of mass ($\dot{m}_{\rm c}\lesssim 0.05\ M_{\odot}$) and redback systems, with heavier companions  \citep[ $ 0.1\ M_{\odot}\ \lesssim \dot{m}_{\rm c} \lesssim 0.5\ M_{\odot}$;][]{r13,CCT+13}. \psr{} is the second black-widow system that was discovered \citep{sbb96}. Its companion is expected to be a $\sim$0.02-0.06 $M_{\odot}$ star, whose exact nature is yet to be determined \citep[see discussions in][]{svbk01,lvt+11}.

Pulsar timing relies on making highly precise measurements of the time at which the radio-beam from a rotating pulsar crosses a radio telescope. These measured times are then compared to a theoretical prediction of these crossing events to derive various properties of the pulsar. A more extensive discussion on pulsar timing and the benefits of MSPs for pulsar timing can be found in \cite{lk05} and other reviews of pulsar timing. 

MSPs are particularly well-suited for this because of their inherent stability and short rotation periods. Even though the pulsars in black-widow systems are MSPs, they are typically excluded from high-precision pulsar timing experiments since several of them have been observed to display variability in their orbital parameters, in particular the orbital period. This variability may be due to many reasons like the interaction of the pulsar with the companion, the presence of excess gas around the companion's orbit or the companion's mass loss.

However, only a limited number of studies so far have tried to identify if the variability of such pulsars can be modelled by introducing new parameters into the pre-existing timing models or by defining new timing models for such systems. Given the recent increase in the number of MSPs detected, in large part from surveys of Fermi-LAT sources \citep{2FGLCat}, and the rapid growth in the sensitivity and bandwidth of modern digital receiver systems for pulsar timing making it possible to detect variations in much greater detail, it is pertinent to address this long-standing question. 

\psr{} has been continuously timed since its discovery in 1995 \citep{sbb96} and therefore the dataset presented in the following analysis represents the longest timing baseline currently published for eclipsing black-widow systems. Given this long time-baseline and other favourable properties discussed in the following sections, this dataset offers an ideal opportunity to attempt such an exercise. 

Previous pulsar timing analyses of \psr{} have shown that the orbital period, $P_{\rm{b}}$, and projected semi-major axis, $x$, undergo secular variations \citep{sbm+98, dlk+01, lvt+11}. These variations are possibly linked to the variations of the gravitational quadrupole moment of the companion and induced by variations of the mass quadrupole of the companion as its oblateness varies due to rotational effects \citep{lvt+11}. These variations may arise due to a differential rotation of the outer layers of the companion \citep{as94} or due to variations in the activity of the magnetic field of the companion as in the \cite{lr01} model. Similar variations have been measured for a few other pulsars in BW systems like PSR J1959+2048 \citep[PSR B1957+20;][]{fst88}, PSRs J0024$-$7204J and PSRs J0024$-$7204O \citep[47 Tuc J and O;][]{fck+03}, PSR J1807$-$2459A \citep[NGC 6544A;][]{LFR+12} and PSR J1731$-$1847 \citep{NBB+14}.

The binary system containing \psr{} has also been recently detected in Fermi-LAT and {\it Chandra}/ACIS data \citep{wkh+12}. The $\gamma$-ray luminosity is $7.66\times10^{32}\  \rm{erg\ s^{-1}}$. The inferred spin-down power, ${\dot{E}}$, from radio observations is $\sim 5.49 \times 10^{33}\ \rm{erg\ s^{-1}}$. The $\gamma$-ray luminosity, therefore, represents $\sim 15\%$ of the total spin-down power, which is consistent with other MSPs for which such a detection has been made. The $\gamma$-ray emission from the system appears to be well fit by a model of emission in the `outer gap accelerator', as discussed in \cite{tct12}. Using the new ephemerides presented here, it may be possible to detect the orbital dependence of pulsed emission from \psr{}.

The $X$-ray luminosity is $1.01\times10^{30}\ \rm{erg\ s^{-1}}$ \citep{wkh+12} and the data do not present any evidence for bursts, which suggests that the companion is stable and does not undergo sudden deformations. The flux values fit well for a model with emission from the intra-binary shock, the polar caps and synchrotron emission from the pulsar magnetosphere \citep{wkh+12}.

This work provides an update on the timing of \psr{} and presents an improved analysis. Two complementary timing models for \psr{} are provided, one capable of handling small eccentricities and another, utilising orbital-frequency derivatives. A new method for measuring the variations in the orbital period, $\Delta P_{\rm{b}}$, by measuring the change in the epoch of ascending node, $T_{\rm{asc}}$ is also presented. 

\section{Observations and Data Analysis}\label{sec:obs_analysis}
The bulk of the dataset used for the timing analysis consists of pulse times-of-arrival (henceforth; ToAs) derived from data from four European Pulsar Timing Array (EPTA) telescopes\footnote{These are the Effelsberg 100-m radio telescope, the Lovell radio telescope at Jodrell Bank, the Nan\c{c}ay radio telescope and the Westerbork Synthesis Radio Telescope. A fifth telescope; the Sardinia Radio Telescope (SRT), has just entered its initial operational phase and therefore no data from the SRT are included here.} and extend from 2003 to 2015. To extend the analysis and to test for consistency with previous analyses, ToAs \citep[obtained from the][dataset]{lvt+11} from the EPTA telescopes, in the period 1995 to 2009, and the Parkes radio telescope, extending from 1995 to 1998, were added to the dataset. Wherever possible, these ToAs were replaced with new ToAs derived from data processed as described later in this section.

As a result of the extended temporal coverage, data-files (henceforth, archives) from a number of pulsar data recording instruments or `backends' are included in the dataset. These include the Effelsberg-Berkeley Pulsar Processor (EBPP), the Berkeley-Orleans-Nan\c{c}ay (BON) instrument, the Digital Filter Bank (DFB) and the Pulsar~Machine~I (PuMa-I) backend, all described in \citep{DCL+16} as well as the Analogue Filter Bank (AFB) \citep{sl96} at Jodrell Bank and, the new generation of pulsar timing backends, namely, PuMa-II at the WSRT \citep{KSS08}, PSRIX at Efflesberg \citep{LKG+16}, ROACH at Jodrell Bank \citep{BJK+16} and the Nan\c{c}ay Ultimate Pulsar Processing Instrument (NUPPI) at Nan\c{c}ay \citep{dbc+11}. The names of all the backends and their respective telescopes can be found in \autoref{tab:rcvr}.

The archives from all the backends were first re-weighted by the sqaure root of the signal-to-noise ratio (S/N) and then grouped into 5-minute integrations using the {\tt psradd} tool from the {\sc PSRchive}\footnote{Commit hash  - 87357c2; \href{URL}{psrchive.sourceforge.net}} suite \citep{hvm04,sdo12}.

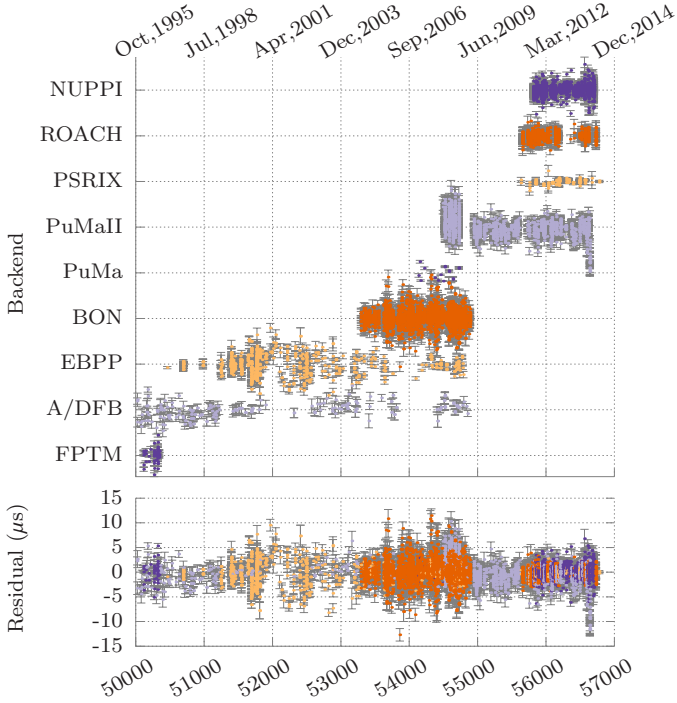
\begin{figure}
\vspace*{6pt}
{\small \input{plots/epslatex/BTOA.tex}}
\vspace*{-20pt}
\caption{Plot of ToAs as a function of MJD. The bottom plot shows the timing residual from fitting the ToAs to the BTX model (see \protect \autoref{sec:obs_analysis}). The top plot shows the same but with manually introduced offsets to show the ToAs grouped by their respective backends. See \protect \autoref{tab:rcvr} for the details of the backends.}
\label{fig:toas}
\end{figure}

ToAs were generated via cross-correlations of the time-integrated, frequency-scrunched, total intensity profiles with noise-free analytical templates, constructed by fitting high S/N pulse profiles with a set of von Mises functions using the {\tt paas} tool of {\sc PSRchive}. These templates were manually aligned using {\tt pas}. The {\tt pat} tool from the same suite was used to generate ToAs, with the Fourier Domain with Markov-chain Monte-Carlo (FDM) algorithm \citep[a re-implementation of][]{Taylor92} and goodness-of-fit (GOF) flags were enabled for the ToAs, as advised by \cite{VLH+16}. A summary of the data from the different backends and telescopes is provided in \autoref{tab:rcvr}. \autoref{fig:toas} shows a plot of the timing residuals for the entire 21-year span, when the ToAs are fitted to the BTX model, as explained below.

Instrumental offsets between the various backends were corrected for by using `JUMP' statements, which allow correct error-propogation.

\input{tables/rcvr.tex}

For the PSRIX backend \citep{LKG+16} at the Effelsberg radio-telescope, which has a total bandwidth of 200~MHz at 21-cm wavelength and the archives with the highest S/N (up to $\sim$4000, for a particular observation), archives were tested for frequency evolution of the pulse shapes. The data were split into 25~MHz channels and analytical templates were generated for each band, as explained above. These templates were manually compared using the {\tt paas} tool. No significant differences were detected and the data were recombined into the full 200~MHz band. For the other backends such an exercise is not possible since either the S/N is typically worse or the bandwidth is too low to detect any obvious frequency evolution in the pulse profile. ToAs were also generated by using templates from different backends to test for pulse shape differences between backends. The timing analysis was then carried out using the \textsc{Tempo2}\footnote{version - 2013.9.1 with updated clock files; \href{URL}{www.atnf.csiro.au/research/pulsar/tempo2}} package \citep{hem06}. Observations which were linked to ToAs with unexplained residual offsets $\ga 3\sigma$ were manually investigated. In some cases, manual RFI excision was sufficient to remove the offset. A few ToAs were found to be linked to observations with previously determined time offsets, which were corrected for using the \textsc{Tempo2} TIME keyword in the relevant sections of the ToA files. In a few cases ToAs were found to have offsets which could not be corrected by either of the two methods. In most cases these ToAs were found to have poor GOF values ($\geq$2) from the template matching and therefore, removed from the dataset. These TOAs are being investigated further to determine their possible association with micro-eclispes of the kind demonstrated by \cite{asr+09}. However, their exclusion does not affect any of the conclusions in this analysis.

Similar to previous analyses, ToAs corresponding to orbital phases 0.2 to 0.35 ( determined using the ephemeris presented in \cite{lvt+11} ) were removed as the eclipse region lies within that range. When carrying out a weighted fit, ToAs with large uncertainties contribute only weakly to the timing solutions and can often be discarded without greatly affecting the results. For MJD ranges with dense temporal sampling, a cut-off of 20 $\mu$s was applied. For the MJD range $\sim$52000 - 53000, where the number of ToAs was very low even before a cut-off was applied, only ToAs with uncertainties greater than 60 $\mu$s were removed.

After the ToA selection procedure described above, the ToAs were split into $\sim$1095 day (or 3 year) long `aeons' with an overlap of 365 days between successive aeons. For each aeon the ToAs were fitted to the ELL1 \citep{lcw+01} timing model, while keeping the dispersion measure (DM) fixed and setting the reference epoch to the centre of the aeon. The timing solutions were derived using the NASA-JPL DE421 planetary ephemeris \citep{fwb09}. The reference clock used was the Terrestrial Time standard derived from the `Temps Atomique International' time standard, denoted by TT(TAI) and the final ToAs were corrected according to the Bureau International des Poids et Mesures (BIPM) standards \citep[see e.g.][and references therein]{hem06}. The default \textsc{Tempo2} assumptions for the Solar-wind model were retained for this analysis.

When using data from multiple instruments, it is necessary to correct the possible mis-estimation of the uncertainty of the ToAs in order to correct for the relative weighting of data from different backends. \textsc{Tempo2} error scaling factors (or T2EFACs) were calculated for each backend by applying the timing model derived in the previous step (without re-fitting) and then taking the square root of the reduced $\rm{\chi} ^{2}$. The corresponding ToA uncertainties were then multiplied by these T2EFAC values.

For the ELL1 model the $\rm{\sigma/\sqrt{N}}$ statistic, where $\sigma$ is the timing residual and $N$ is the number of ToAs is used to select the aeon with the most information. From \autoref{tab:aeonwise} this is identified as the epoch starting at MJD 55121. The timing parameters for this aeon are presented in \autoref{tab:ephem} and a comparison with published literature is provided in \autoref{tab:compare}.

\input{tables/aeonwise.tex}

As is obvious from the preceding discussions, the ELL1 model requires updating at regular intervals or aeons. This is a consequence of the orbital variability of this system, as discussed in \autoref{ssec:secvar}. Therefore, the BTX model was used to construct a single timing model encompassing the entire 21 year period.

\input{tables/compare.tex}

The BTX model is a re-implementation of the BT model \citep{bt76} and incorporates higher order derivatives of the orbital-frequency. This model is completely phenomenological and thus, has no predictive power. The model also demands judicious usage since the highest order orbital-frequency derivatives can easily introduce correlations with proper motion components, DM variations and instrumental offsets, particularly in this highly heterogeneous dataset. Eccentricity measurements from the ELL1 models show large variability along with low measurement significance, indicating that these measurements are probably unreliable. Hence, the BTX model was created with eccentricity set to zero. 

To limit the number of orbital-frequency derivatives (OFDs) employed in the BTX model, the reduced $\rm{\chi} ^{2}$ was used as the primary selection criterion. The reduced $\chi^{2}$ remains well above ten until the 17$^{th}$ OFD is introduced. Subsequent OFDs do not affect the reduced $\rm{\chi} ^{2}$ and are not determined with any significance by \textsc{Tempo2}. Amongst the timing models with 13 or more OFDs, the Akaike Information Criterion \citep{Akaike74} also favours the model with 17 OFDs. The BTX timing parameters with 17 OFDs for \psr{} are  presented in \autoref{tab:ephem}.

\input{tables/merged.tex}

The timing models and ToAs are available under `additional online material' at the EPTA web page, accessible via \href{URL}{http://www.epta.eu.org/aom.html}.

\section{Timing Results}\label{sec:timing}
\subsection{Proper motion}\label{ssec:pm}
\psr{} has a low ecliptic latitude of $\sim 8\degr.85$. Typically for such low latitudes, the determination of position is relatively poor \citep{lk05}. Therefore, the resulting measurement of proper motion in declination or ecliptic latitude (depending on the coordinate system used) is imprecise. This is evident in the published values of proper motion in declination, $\mu_{\delta}$, presented in \autoref{tab:compare}.

\begin{figure}
\vspace*{6pt}
{\small \input{plots/epslatex/RA_DEC_1095.tex}}
\vspace*{-20pt}
\caption{Measured values of RA (left) and DEC (right) of \psr{} for each aeon (purple +) and linear fits to those. Black arrows indicate the values at the reference epoch at which the two timing models of \autoref{tab:ephem} are defined, MJD 55655. The fit to the position at the median MJD of each aeon (the finely-dotted orange line) returns $\mu_{\alpha}\ =\ 5.63\pm 0.10\ \rm{mas\ yr^{-1}}$ and  $\mu_{\delta}\ =\ 2.34\pm 0.28\ \rm{mas\ yr^{-1}}$ while the dashed, lilac line represents the values obtained from the BTX model, shown in \autoref{tab:ephem}.}
\label{fig:fitspm}
\end{figure}
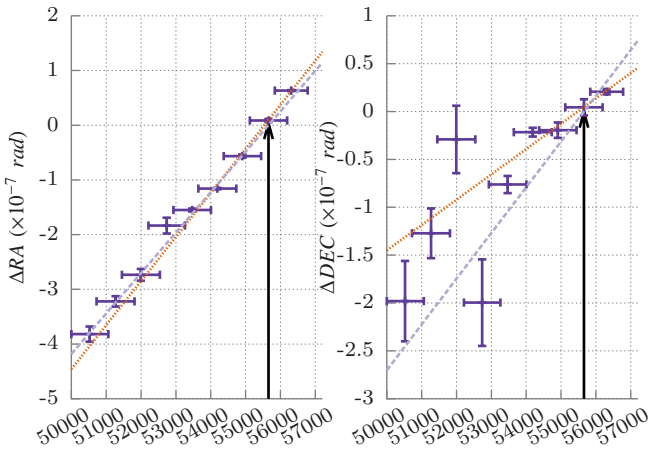

To improve the measurement and utilise the entire 21-year span of the dataset, the measured value of right ascension (RA) and declination (DEC) for each aeon were fitted with a linear function to obtain a mean proper motion. This results in a significant measurement of $\mu_{\alpha}$ and  $\mu_{\delta}$, as shown in \autoref{fig:fitspm} and \autoref{tab:compare}. The fitted values of $\mu_{\alpha}$ and  $\mu_{\delta}$ are inserted into the ELL1 models for each aeon and those models are refitted for the other parameters.

Using an estimated distance of $\simeq$1040 pc \citep[from the NE2001 model of free-electron distribution in the Galaxy,][]{cl03} and a total proper motion, ${\mu_{t} = \sqrt{\mu_{\alpha}^2 + \mu_{\delta}^2} = 6.1\ \pm \ 0.1\ \rm{mas\ yr^{-1}}}$; a 2-D transverse velocity of ${\nu_t = 30\ \pm \ 9\ \rm{km\ s^{-1}}}$ was calculated. This assumes an uncertainty of 30\%\footnote{See \cite{DCL+16} for a discussion on the possible underestimation of uncertainties of the DM derived distances.} in the DM-derived distance mentioned above. The measurement is in agreement with the value of $30\ \pm \ 20\ \rm{km\ s^{-1}}$ measured by \cite{sbm+98} and represents a two-fold increase in precision, even though the uncertainty of the DM-derived distance is assumed to be much greater. It should be noted that this is significantly lower than the average value of $93 \pm 13\ \rm{km\ s^{-1}}$ reported in \cite{DCL+16} for the transverse velocities of binary MSPs. However, it agrees well with the value of $56 \pm 3\ \rm{km\ s^{-1}}$ reported for the binary MSPs with distance measurements from parallaxes.

The proper motion values obtained from the BTX model appear to be inconsistent with those obtained from fitting to position measurements for every aeon using the respective ELL1 models. This is because the proper motion terms and the orbital-frequency derivatives are strongly covariant and therefore the uncertainties in the values obtained from the BTX model are heavily underestimated, reinforcing the need for cautious usage of this model.


\subsection{DM variations}\label{ssec:dmvar}
Since the DM is a measure of the density of the ionised interstellar medium (IISM) along the line of sight to the pulsar, both the motion of the pulsar and the dynamical evolution of the IISM affect this value. While it is possible to obtain the DM from timing, `JUMPS' or instrumental offsets introduced to align the ToAs from the different backends are fully covariant with the DM and prevent an accurate measurement directly from the dataset presented above. Therefore, a DM value of 20.7299(17) $\rm{cm^{-3}pc}$ is adopted from the LOFAR measurements of \cite{kvh+16}.

When simultaneous dual (or multi) frequency observations are available, it is, however, possible to accurately estimate the {\em variation} in the DM. The WSRT Puma-II backend provides observations centred at 345 and 1380~MHz, with a cadence of roughly three weeks. Observations between the two frequencies are sequential, which are separated by, at most, a few days and available for the MJD range $\sim$54600 -- 56800. Since low-frequency observations are more sensitive to the DM variations, these are utilised to measure them instead of the two frequency-band observations of the PSRIX backend. 

To measure DM variations, the PuMa-II ToAs were fitted for DM using the ELL1 model presented in \autoref{tab:ephem}. The ToAs were then split into 100-day long intervals, to ensure enough data were available for a reliable estimate. Each 100-day interval was then refitted for the DM, $P_{\rm{b}}$ and $T_{\rm{0}}$. The fit for $P_{\rm{b}}$ is necessary to ensure that orbital-phase dependent effects do not contaminate the DM measurement, since the observations at 345~MHz and 1380~MHz do not necessarily coincide in orbital phase. 

\begin{figure}
\vspace*{6pt}
\resizebox{\columnwidth}{!}{\small \input{plots/epslatex/dmvar.tex}}
\vspace*{-10pt}
\caption{DM variation from consecutive 327~MHz and 1380~MHz observations with the WSRT which extend over the period 54600-56800. A linear fit (lilac, dashed) and a quadratic fit (orange, finely-dotted) are also shown.}
\label{fig:dm_simul_wsrt}
\end{figure}
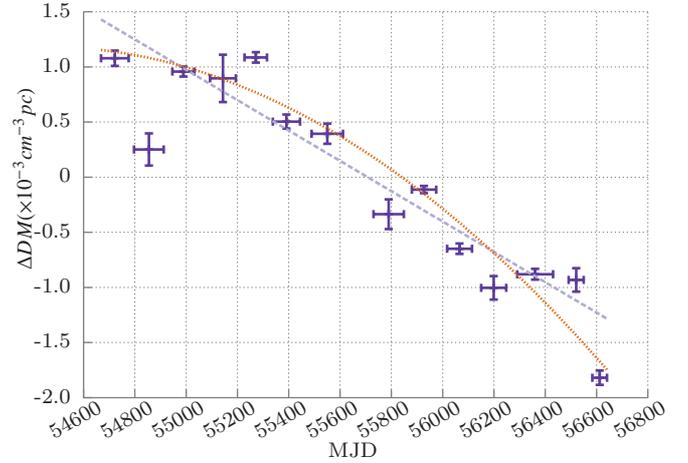

This leads to a significant detection of a DM trend after MJD~54600, which is plotted in \autoref{fig:dm_simul_wsrt}. A quadratic fit returns a reduced-$\chi^{2}$ of 3.5 while a linear fit performs not much worse, with a reduced-$\chi^{2}$ of 6. The linear trend appears to show a weakly sinusoidal residual, with a `best-fit' period of $\sim$940 days but this residual becomes insignificant with the quadratic model and therefore, no higher-order model was considered.

While it is quite possible that such variations may be present before MJD 54600, the lack of sensitivity due to sparse and inhomogeneous multi-frequency observations lead to typical DM measurement uncertainties of $\sim 1 \times 10^{-3}$ to about $\sim 3 \times 10^{-4}\ \rm{cm^{-3}pc}$. These uncertainties, which may well be severely underestimated, prevent any firm conclusion on the DM evolution. Furthermore, because no combination of two observing systems at different frequencies is continuously present before MJD 54600, any effort to measure DM variations in that MJD range is necessarily corrupted by the arbitrary phase offsets used to align the data from different instruments. The WSRT data which provide continuous data at two frequencies after MJD 54600 provide a DM precision of $\la\ 3\ \times\ 10^{-4}\ \rm{cm^{-3}pc}$ over 100-day intervals, allowing accurate DM modeling over that period. 

Traditionally, wherever a DM trend is observed, it is corrected for by introducing DM derivatives.\footnote{For detailed reviews on modern DM correction methods see \cite{VLH+16}, \cite{dfg+13} or \cite{LTM+15}.} Given the large uncertainties in the earliest eras and to prevent over-fitting or accidentally introducing excess white noise in the timing, only those ToAs belonging to the period over which a clear DM trend is measured are corrected for the DM trend modelled by the quadratic fit shown in \autoref{fig:dm_simul_wsrt}. This is implemented by introducing a \textsc{Tempo2} DM offset flag (-dmo) for the ToAs lying in the MJD range 54600-56800.

\subsection{Secular variations}\label{ssec:secvar}
\begin{figure}
\resizebox{\columnwidth}{!}{\small \input{plots/epslatex/year.tex}}
\vspace*{-23pt}
\caption{Change in $P_{\rm{b}}$ and $x$ measured by fitting for $P_{\rm{b}}$, $x$ and $T_{\rm{asc}}$ for eras of length 365 days with an overlap of 30 days, where possible.}
\label{fig:pbxvar}
\vspace*{-18pt}
\end{figure}
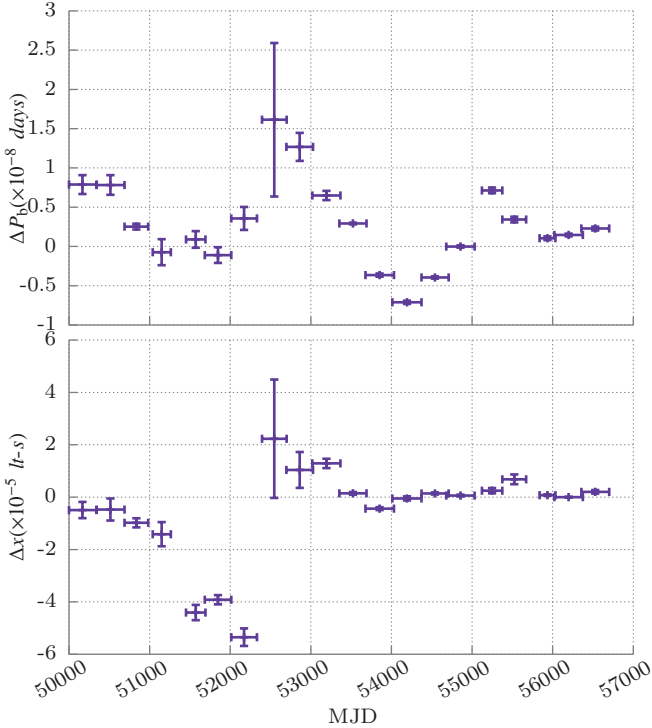

Following \cite{lvt+11}, variations in the binary period ($P_{\rm{b}}$) and the projected semi-major axis ($x$) were measured by splitting the ToAs into `eras' of approximately 365 days. The results of reproducing and extending the Lazaridis analysis\footnote{In the \cite{lvt+11} analysis, timing models are first derived for the largest MJD range over which a \textsc{Tempo2} fit converges, which are analogous to `aeons' in the present work. Then, the variations in $P_{\rm{b}}$ and $x$ are measured by fitting for $P_{\rm{b}}$, $x$ and $T_{\rm{asc}}$ simultaneously for 300-day periods with an overlap of 30 days.} are presented in \autoref{fig:pbxvar}.

The simultaneous fitting of  $P_{\rm{b}}$, $x$ and $T_{\rm{asc}}$, as in \cite{lvt+11}, is undesirable since $P_{\rm{b}}$ and $T_{\rm{asc}}$ are fully covariant parameters. In practice, wherever good orbital phase coverage ($\geq$ 60\%) is available, the measurement of $T_{\rm{asc}}$ is far more accurate and reliable since it measures the orbital phase and requires less information for its calculation than $P_{\rm{b}}$. Due to the high cadence and long durations of the Nan\c{c}ay, Jodrell Bank and WSRT observations and full orbital observations at Effelsberg, especially in the latest years, it is possible to carry out such a measurement with much greater precision than was previously attempted.

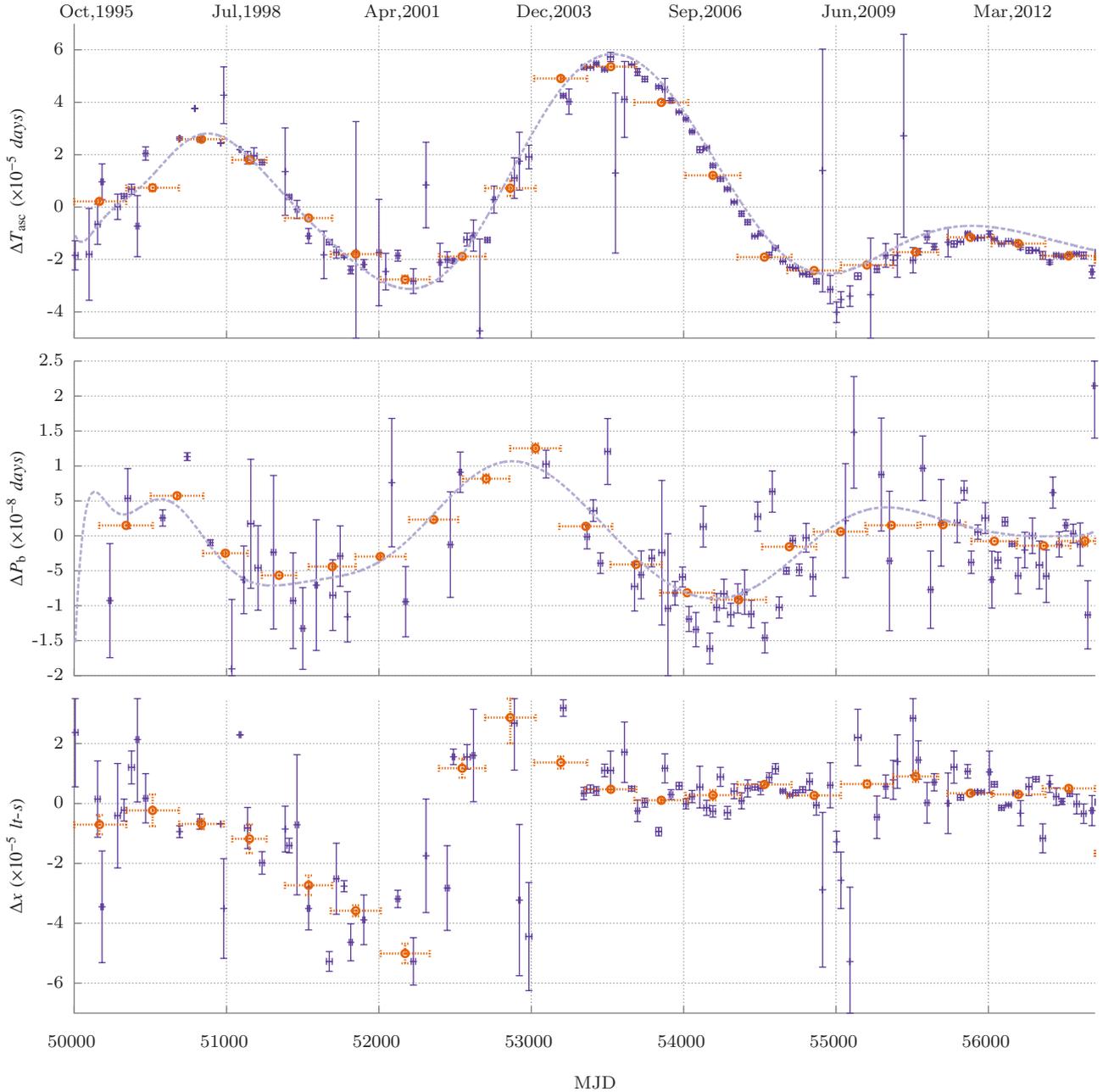
\begin{figure*}
{\small \input{plots/epslatex/tascyear.tex}}
\vspace*{-43pt}
\caption{Plot of $\Delta T_{\rm{asc}}$, $\Delta P_{\rm{b}}$ and $\Delta x$ measured from fitting for $x$ and $T_{\rm{asc}}$ only for epochs with a length of 45 (purple +) and 365 (orange $\odot$) days, along with the variations described by the BTX model (lilac, dashed ). To improve the readability of the graphs for $\Delta P_{\rm{b}}$ and $\Delta x$, points with uncertainties comparable to the y-range of the graph (typically in the earliest epochs) are removed. The prominent fluctuations for the BTX prediction of $\Delta P_{\rm{b}}$ at $\sim$ MJD 50100-50600 agree with the measured ( but unplotted ) values, as can be discerned from the $\Delta T_{\rm{asc}}$ plot.}
\label{fig:tascyear}
\end{figure*}

By keeping $P_{\rm{b}}$ constant for all eras, and fitting for $T_{\rm{asc}}$, $x$ and the Laplace-Lagrange parameters, ${\eta\ =\ e \cdot\ sin \omega}$ and ${\kappa\ =\ e \cdot\ cos\ \omega}$ simultaneously, the change in $T_{\rm{asc}}$ is measured. 
The change in $P_{\rm{b}}$ measured at time $t_1$, $\Delta P_{{\rm b},t_1}$, is then calculated using the equation
\begin{equation}
  \Delta P_{{\rm b} ,t_1} \ =\ \frac {T_{{\rm asc} ,t_1} - T_{{\rm asc},t_0} }{t_1 - t_0 } \times {P_{{\rm b} ,ref}}
\end{equation}
where $T_{{\rm asc},t_0}$ and $T_{{\rm asc} ,t_1}$ are the values of $T_{\rm{asc}}$ at two neighbouring eras $t_0$ and $t_1$. $P_{{\rm b},ref}$ is a constant $P_{\rm{b}}$ value chosen from the $P_{\rm{b}}$ values for each epoch, such that the measured $\Delta T_{\rm{asc}}$ values do not show any obvious slope. The resulting  $\Delta P_{\rm{b}}$ variations and the $\Delta T_{\rm{asc}}$ from which they are derived are plotted in \autoref{fig:tascyear}, along with the simultaneous $\Delta x$ measurements. The measured values for all three parameters are over-plotted with the interpolation of the change in $\Delta T_{\rm{asc}}$ as obtained from the BTX model \citep[see, e.g.,][]{NBB+14}. The  excellent agreement serves to further confirm the applicability of the BTX model.

Comparing the $P_{\rm{b}}$ variations derived from the $T_{\rm{asc}}$ variations in \autoref{fig:tascyear} and those in \autoref{fig:pbxvar}, derived from the \cite{lvt+11} method, it is apparent that fitting for all three parameters introduces a `smoothing' effect. This is likely due to the covariance of $P_{\rm{b}}$ and $T_{\rm{asc}}$ and thus demonstrates the importance of estimating $\Delta P_{\rm{b}}$ from fitting for $T_{\rm{asc}}$ and $x$ simultaneously. It should be noted that for all the eras that were analysed, $P_{\rm{b}}$ and $T_{\rm{asc}}$ were found to be either strongly correlated or anti-correlated (~$\lvert$corr.$\rvert\geq$0.9~), with a somewhat alternating behaviour, while the $P_{\rm{b}}$ and $x$ are always weakly correlated (~$\lvert$corr.$\rvert\leq$0.3~). Finally, $T_{\rm{asc}}$ and $x$ are always very weakly correlated (~$\lvert$corr.$\rvert\ll$0.3~).

As can be seen from \autoref{fig:tascyear} the new analysis is in qualitative agreement with the measurements presented in \cite{lvt+11} and the system appears to have entered a `quieter' phase. For brevity, only a summary of the maximum possible contribution to the secular variations from the various possible sources is presented in \autoref{tab:annuvar}. For a full discussion of these, see \cite{lvt+11}.

Variations in the orbital period can be attributed to contributions due to gravitational-wave emission ($\dot{P}_{\rm{b}} ^{\rm{GW}}$), changing Doppler shift ($\dot{P}_{\rm{b}}^{\dot{\rm{D}}}$), mass loss from the companion ($\dot{P}_{\rm{b}}^{\dot{m}}$), tidal interactions between the companion and the pulsar ($\dot{P}_{\rm{b}} ^{\rm{T}}$) and variations of the gravitational quadrupole moment of the companion star ($\dot{P}_{\rm{b}} ^{\rm{Q}}$) \citep[see, for instance][]{lk05}\footnote{The sign on the $\dot{P}_{\rm{b}}^{\dot{\rm{D}}}$ and $\dot{x}^{\dot{\rm{D}}}$ terms are made positive for the sake of uniformity here.}:

\begin{equation}\label{eq:Pb_terms}
\dot{P}_{\rm{b}} ^{obs} = \dot{P}_{\rm{b}} ^{\rm{GW}} + \dot{P}_{\rm{b}}^{\dot{\rm{D}}} + \dot{P}_{\rm{b}}^{\dot{m}} + \dot{P}_{\rm{b}}^{\rm{T}} + \dot{P}_{\rm{b}}^{\rm{Q}}
\end{equation}

Similarly, the secular variations in the projected semi-major axis can be split into contributions due to radiation of gravitational waves ($\dot{x}^{\rm{GW}}$), the proper motion of the pulsar ($\dot{x}^{\rm{PM}}$), varying aberrations ($\frac{d\epsilon_{\rm{A}}}{dt}$), changing Doppler shift ($\dot{x}^{\dot{\rm{D}}}$), mass loss in the binary system ($\dot{x}^{\dot{m}}$), variations of the gravitational quadrupole moment of the companion star ($\dot{x}^{\rm{Q}}$), spin-orbit coupling of the companion ($\dot{x}^{\rm{SO}}$) and a second, or planetary, companion ($\dot{x}^{\rm{p}}$).

\begin{equation}\label{eq:x_terms}
\dot{x}^{obs} = \dot{x}^{\rm{GW}} + \dot{x}^{\rm{PM}} + \frac{d\epsilon_{\rm{A}}}{dt} + \dot{x}^{\dot{\rm{D}}} \\
+ \dot{x}^{\dot{m}} + \dot{x}^{\rm{Q}} + \dot{x}^{\rm{SO}} + \dot{x}^{\rm{p}}
\end{equation}

\input{tables/annuvar.tex}

For the observed 21-year baseline, the maximum $\dot{P}_{\rm{b}}$ is $\sim \PBDmax{}$ and the minimum is $\sim \PBDmin{}$. From \autoref{tab:annuvar} it is evident that the first four terms of \autoref{eq:Pb_terms} cannot drive the observed $\Delta P_{\rm{b}}$ variations independently. Therefore, the hypothesis of \cite{lvt+11} that the  mass quadrupole variations in the companion are the most likely drivers of the observed $\Delta P_{\rm{b}}$ variations is recovered.

Similarly, from \autoref{fig:tascyear}, the variation of the projected semi-major axis shows a strong `feature' in the MJD-range $\sim$51000 to 53000, which is not present in the remaining data. Since the correlation between $x$ and $T_{\rm{asc}}$ or $x$ and $P_{\rm{b}}$ is very weak, the differences between the bottom panels of \autoref{fig:pbxvar} and \autoref{fig:tascyear} are marginal, although the uncertainties in the second case are typically smaller for the 365-day epochs. 

As in the case of the $\Delta P_{\rm{b}}$ variations, the terms of \autoref{eq:x_terms} for which values are presented in \autoref{tab:annuvar} are not likely to be independent drivers of the variations in $\Delta x$. This implies that the \cite{lvt+11} conjecture that the classical spin-orbit coupling term combined with the GQ term is the most likely driver for the $\Delta x$ variations is also recovered.

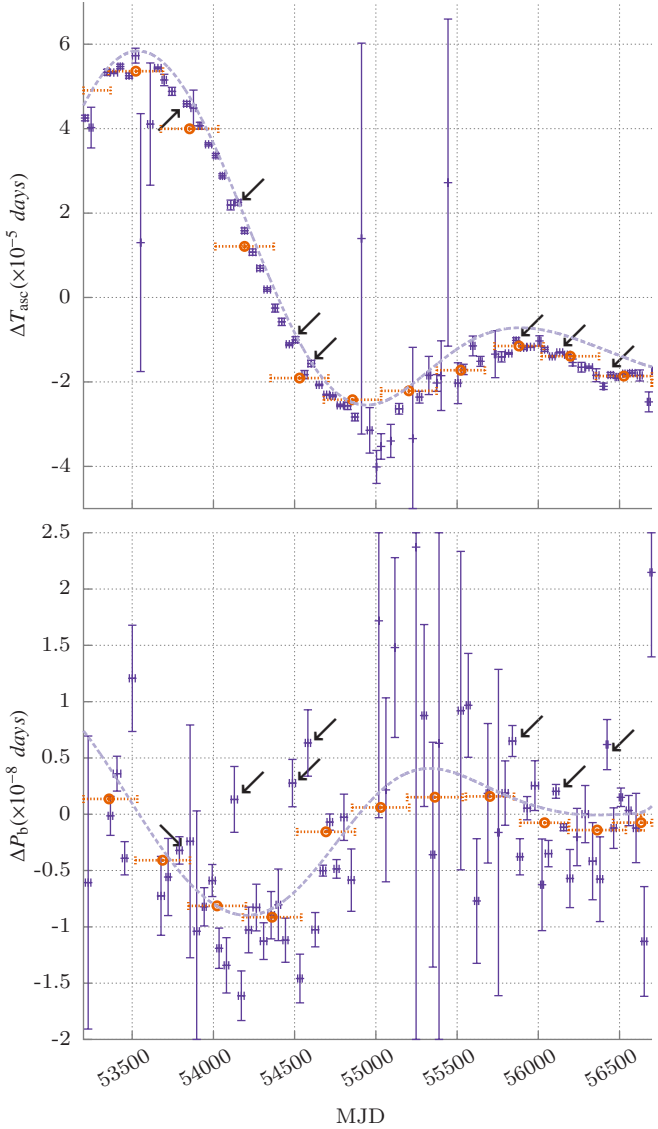
\begin{figure}
{\small \input{plots/epslatex/tascyearzoom2.tex}}
\vspace*{-33pt}
\caption{Zoomed in plot of $\Delta T_{\rm{asc}}$, $\Delta P_{\rm{b}}$ and $\Delta x$ measured from fitting for $x$ and $T_{\rm{asc}}$ only for 45 (purple +) and 365 (orange $\odot$) day long epochs, along with the predicted variations from the BTX model (lilac, dashed). Solid black arrows indicate epochs where the derivative of $T_{\rm{asc}}$ abruptly changes sign.}
\label{fig:tascyearzoom}
\end{figure}

In addition to recovering the long-term fluctuations, the derivation of $\Delta P_{\rm{b}}$ from $\Delta T_{\rm{asc}}$ reveals small-scale variations, as indicated with black arrows in \autoref{fig:tascyearzoom}. These points lie $\ga$4$\sigma$ away from their local means and do have corresponding values with negative offsets. Given the results from \cite{wkh+12} presented in \autoref{sec:intro}, it remains unclear what processes could lead to such deviations. 

It is evident that continued multi-band monitoring of \psr{} is necessary to reveal the origin of these sudden, sharp increases in the orbital period. If these changes are a result of activity of the companion, a greater understanding of the origin of these changes might help to understand the processes which drive state changes in the `transitioning' MSP systems, i.e., binaries where the MSP alternates between accreting and radio-pulsar states \citep[see, e.g.,][]{SAH+14}.

Given the high cadence and regular sampling in the later aeons, a test for the presence of a second companion, possibly of planetary dimensions, is carried out as well. This involves testing for the presence of higher-order derivatives of pulse frequency in the timing solution \citep{jr97}. The extrema of the second and third order frequency derivatives from \textsc{Tempo2} fits to the aeons are $ -4.1(8) \times 10^{-24}\ \rm{s^{-3}}\ \leqslant f^{(2)}_{max} \leqslant  3.0(19) \times 10^{-24}\ \rm{s^{-3}}$ and $ 1.1(6) \times 10^{-30}\ \rm{s^{-4}}\ \leqslant f^{(3)}_{max} \leqslant 2.1(9) \times 10^{-30}\ \rm{s^{-4}}$. Since these values are at best marginally significant and in the absence of any supporting evidence from optical observations, the hypothesis of a second companion to \psr{} remains unjustified. 

\section{High-precision timing prospects}\label{sec:hiprectiming}
Due to the complicated and somewhat arbitrary orbital variability that some pulsars in BW systems have been shown to exhibit \citep[e.g.,][etc.]{nat00,fck+03,LFR+12,NBB+14}, these sources have been traditionally left out of high-precision pulsar-timing campaigns. With the recent increase in the number of BW systems discovered among the FERMI sources \citep{2FGLCat}, it will soon be possible to quantify these instabilities for a larger sample. As a counter-example to the current practice, the pulsar of the BW system J0610$-$2100 has recently been added to the list of sources for the EPTA \citep{DCL+16} and has, so far, provided stable timing.

Simulations for pulsars timed using the BTX model by \cite{BRD+15} show that only a small percentage of the power from gravitational waves is likely to be absorbed into the higher-order orbital-frequency derivatives and again, appear to favour the inclusion of such pulsars in PTAs. However, \cite{BRD+15} do not take into consideration variations of $x$, as identified for the BW system J2051$-$0827.

The timing analysis presented here demonstrates the practical usability of the BTX model for such systems. However, it should also be noted that the GOF for the BTX model is still rather low as some variations remain unaccounted for.

It is probably an opportune coincidence that the BW system J2051$-$0827 has entered a relatively stable phase, suggesting greater usability for a PTA. Even without addressing some of the ambiguities in the fundamental properties of this system, for both the ELL1 and BTX models, the present analysis shows it is possible to obtain timing residuals of the order of $\sim$5.0 $\rm{\mu s}$, quite comparable to the timing precision of several sources already in the PTAs \citep{VLH+16}. In the intermediate-to-high S/N regime of gravitational wave background observations, where the number of pulsars becomes more important than very high timing precision \citep{SEJ+13} timing residuals of the order of $ 1 \mu s$ could be sufficient. With the advent of the new `ultra-broadband' backends (Karuppusamy, private communication) and rapid increases in sensitivity, this does not appear to be an unrealistic goal. 

\section{Summary}\label{sec:summary}
A timing update on \psr{} is presented, along with timing models for the BTX and ELL1 models of \textsc{Tempo2}. An improved estimate of the mean proper motion is also made, giving a value of $30\pm 9\ \rm{km\ s^{-1}}$. A significant decrease in the DM of $\sim 2.5\times 10^{-3} \rm{cm^{-3}pc}$ is detected for the MJD range 54600-56800 and corrections are incorporated in the ToA file.

A more robust analysis is performed by reducing covariant terms and it is shown that the resulting measurements are more precise and consistent with earlier analyses. The variations of the orbital period are detected over more than a full `period', supporting earlier analyses that suggested that these variations arise from cyclic variations in the companion, instead of a tertiary star or planet. In addition, small-scale fluctuations in the $P_{\rm{b}}$ variations are detected.

The continued timing of \psr{} shows that the variation of the projected semi-major axis appears to have decreased and does not show the extreme behaviour observed at an earlier epoch, lending hope that the black widow system containing \psr{} may be included in PTAs in the near future.

\section*{Acknowledgements}\label{sec:ack}
The authors acknowledge the support of colleagues in the European Pulsar Timing Array {(EPTA:~\href{URL}{http://www.epta.eu.org})}. The  EPTA is a collaboration of European institutes working towards the direct detection of low-frequency gravitational waves and the implementation of the Large European Array for Pulsars (LEAP).

Part of this work is based on observations with the 100-m telescope of the Max-Planck-Institut f\"{u}r Radioastronomie (MPIfR) at Effelsberg.

Access to the Lovell Telescope is supported through an STFC consolidated grant.

The Nan\c{c}ay radio telescope is part of the Paris Observatory, associated with the Centre National de la Recherche Scientifique (CNRS), and partially supported by the R\'{e}gion Centre in France.

The Westerbork Synthesis Radio Telescope is operated by the Netherlands Institute for Radio Astronomy (ASTRON) with support from The Netherlands Foundation for Scientific Research NWO.

The Parkes radio telescope is a part of the Australian Telescope National Facility (ATNF) which is funded by the Commonwealth of Australia for operation as a national facility managed by the CSIRO.

The authors are also extremely grateful to the past and present operators and other staff at the respective radio telescopes, without whose cheerful and unflagging support much of the observations presented here would have been impossible. 

RNC acknowledges the support of the International Max Planck Research School Bonn/Cologne and the {Bonn-Cologne} Graduate School. PL acknowledges the support of the International Max Planck Research School Bonn/Cologne. SO is supported by the Alexander von Humboldt foundation. CGB acknowledges support from the European Research Council under the European Union's Seventh Framework Programme (FP/2007-2013) / ERC Grant Agreement nr. 337062 (DRAGNET; PI Jason Hessels).




\bibliographystyle{mnras}
\bibliography{journals,PSRJ2051_v3} 



\appendix


\bsp
\label{lastpage}
\end{document}

%% file: plots/epslatex/BTOA.tex
\begingroup
  \makeatletter
  \providecommand\color[2][]{%
    \GenericError{(gnuplot) \space\space\space\@spaces}{%
      Package color not loaded in conjunction with
      terminal option `colourtext'%
    }{See the gnuplot documentation for explanation.%
    }{Either use 'blacktext' in gnuplot or load the package
      color.sty in LaTeX.}%
    \renewcommand\color[2][]{}%
  }%
  \providecommand\includegraphics[2][]{%
    \GenericError{(gnuplot) \space\space\space\@spaces}{%
      Package graphicx or graphics not loaded%
    }{See the gnuplot documentation for explanation.%
    }{The gnuplot epslatex terminal needs graphicx.sty or graphics.sty.}%
    \renewcommand\includegraphics[2][]{}%
  }%
  \providecommand\rotatebox[2]{#2}%
  \@ifundefined{ifGPcolor}{%
    \newif\ifGPcolor
    \GPcolortrue
  }{}%
  \@ifundefined{ifGPblacktext}{%
    \newif\ifGPblacktext
    \GPblacktexttrue
  }{}%
  \let\gplgaddtomacro\g@addto@macro
  \gdef\gplbacktext{}%
  \gdef\gplfronttext{}%
  \makeatother
  \ifGPblacktext
    \def\colorrgb#1{}%
    \def\colorgray#1{}%
  \else
    \ifGPcolor
      \def\colorrgb#1{\color[rgb]{#1}}%
      \def\colorgray#1{\color[gray]{#1}}%
      \expandafter\def\csname LTw\endcsname{\color{white}}%
      \expandafter\def\csname LTb\endcsname{\color{black}}%
      \expandafter\def\csname LTa\endcsname{\color{black}}%
      \expandafter\def\csname LT0\endcsname{\color[rgb]{1,0,0}}%
      \expandafter\def\csname LT1\endcsname{\color[rgb]{0,1,0}}%
      \expandafter\def\csname LT2\endcsname{\color[rgb]{0,0,1}}%
      \expandafter\def\csname LT3\endcsname{\color[rgb]{1,0,1}}%
      \expandafter\def\csname LT4\endcsname{\color[rgb]{0,1,1}}%
      \expandafter\def\csname LT5\endcsname{\color[rgb]{1,1,0}}%
      \expandafter\def\csname LT6\endcsname{\color[rgb]{0,0,0}}%
      \expandafter\def\csname LT7\endcsname{\color[rgb]{1,0.3,0}}%
      \expandafter\def\csname LT8\endcsname{\color[rgb]{0.5,0.5,0.5}}%
    \else
      \def\colorrgb#1{\color{black}}%
      \def\colorgray#1{\color[gray]{#1}}%
      \expandafter\def\csname LTw\endcsname{\color{white}}%
      \expandafter\def\csname LTb\endcsname{\color{black}}%
      \expandafter\def\csname LTa\endcsname{\color{black}}%
      \expandafter\def\csname LT0\endcsname{\color{black}}%
      \expandafter\def\csname LT1\endcsname{\color{black}}%
      \expandafter\def\csname LT2\endcsname{\color{black}}%
      \expandafter\def\csname LT3\endcsname{\color{black}}%
      \expandafter\def\csname LT4\endcsname{\color{black}}%
      \expandafter\def\csname LT5\endcsname{\color{black}}%
      \expandafter\def\csname LT6\endcsname{\color{black}}%
      \expandafter\def\csname LT7\endcsname{\color{black}}%
      \expandafter\def\csname LT8\endcsname{\color{black}}%
    \fi
  \fi
  \setlength{\unitlength}{0.0500bp}%
  \begin{picture}(4818.00,5102.00)%
    \gplgaddtomacro\gplbacktext{%
      \colorrgb{0.50,0.50,0.50}%
      \put(864,1914){\makebox(0,0)[r]{\strut{}FPTM}}%
      \colorrgb{0.50,0.50,0.50}%
      \put(864,2258){\makebox(0,0)[r]{\strut{}A/DFB}}%
      \colorrgb{0.50,0.50,0.50}%
      \put(864,2601){\makebox(0,0)[r]{\strut{}EBPP}}%
      \colorrgb{0.50,0.50,0.50}%
      \put(864,2944){\makebox(0,0)[r]{\strut{}BON}}%
      \colorrgb{0.50,0.50,0.50}%
      \put(864,3287){\makebox(0,0)[r]{\strut{}PuMa}}%
      \colorrgb{0.50,0.50,0.50}%
      \put(864,3631){\makebox(0,0)[r]{\strut{}PuMaII}}%
      \colorrgb{0.50,0.50,0.50}%
      \put(864,3974){\makebox(0,0)[r]{\strut{}PSRIX}}%
      \colorrgb{0.50,0.50,0.50}%
      \put(864,4317){\makebox(0,0)[r]{\strut{}ROACH}}%
      \colorrgb{0.50,0.50,0.50}%
      \put(864,4661){\makebox(0,0)[r]{\strut{}NUPPI}}%
      \colorrgb{0.50,0.50,0.50}%
      \put(772,4900){\rotatebox{30}{\makebox(0,0)[l]{\strut{} Oct,1995}}}%
      \colorrgb{0.50,0.50,0.50}%
      \put(1282,4900){\rotatebox{30}{\makebox(0,0)[l]{\strut{} Jul,1998}}}%
      \colorrgb{0.50,0.50,0.50}%
      \put(1792,4900){\rotatebox{30}{\makebox(0,0)[l]{\strut{} Apr,2001}}}%
      \colorrgb{0.50,0.50,0.50}%
      \put(2302,4900){\rotatebox{30}{\makebox(0,0)[l]{\strut{} Dec,2003}}}%
      \colorrgb{0.50,0.50,0.50}%
      \put(2811,4900){\rotatebox{30}{\makebox(0,0)[l]{\strut{} Sep,2006}}}%
      \colorrgb{0.50,0.50,0.50}%
      \put(3321,4900){\rotatebox{30}{\makebox(0,0)[l]{\strut{} Jun,2009}}}%
      \colorrgb{0.50,0.50,0.50}%
      \put(3831,4900){\rotatebox{30}{\makebox(0,0)[l]{\strut{} Mar,2012}}}%
      \colorrgb{0.50,0.50,0.50}%
      \put(4341,4900){\rotatebox{30}{\makebox(0,0)[l]{\strut{} Dec,2014}}}%
      \put(80,3287){\rotatebox{-270}{\makebox(0,0){\strut{}Backend}}}%
    }%
    \gplgaddtomacro\gplfronttext{%
    }%
    \gplgaddtomacro\gplbacktext{%
      \colorrgb{0.50,0.50,0.50}%
      \put(864,480){\makebox(0,0)[r]{\strut{}-15}}%
      \colorrgb{0.50,0.50,0.50}%
      \put(864,666){\makebox(0,0)[r]{\strut{}-10}}%
      \colorrgb{0.50,0.50,0.50}%
      \put(864,851){\makebox(0,0)[r]{\strut{}-5}}%
      \colorrgb{0.50,0.50,0.50}%
      \put(864,1037){\makebox(0,0)[r]{\strut{} 0}}%
      \colorrgb{0.50,0.50,0.50}%
      \put(864,1222){\makebox(0,0)[r]{\strut{} 5}}%
      \colorrgb{0.50,0.50,0.50}%
      \put(864,1408){\makebox(0,0)[r]{\strut{} 10}}%
      \colorrgb{0.50,0.50,0.50}%
      \put(864,1593){\makebox(0,0)[r]{\strut{} 15}}%
      \colorrgb{0.50,0.50,0.50}%
      \put(672,100){\rotatebox{30}{\makebox(0,0)[l]{\strut{} 50000}}}%
      \colorrgb{0.50,0.50,0.50}%
      \put(1182,100){\rotatebox{30}{\makebox(0,0)[l]{\strut{} 51000}}}%
      \colorrgb{0.50,0.50,0.50}%
      \put(1692,100){\rotatebox{30}{\makebox(0,0)[l]{\strut{} 52000}}}%
      \colorrgb{0.50,0.50,0.50}%
      \put(2202,100){\rotatebox{30}{\makebox(0,0)[l]{\strut{} 53000}}}%
      \colorrgb{0.50,0.50,0.50}%
      \put(2711,100){\rotatebox{30}{\makebox(0,0)[l]{\strut{} 54000}}}%
      \colorrgb{0.50,0.50,0.50}%
      \put(3221,100){\rotatebox{30}{\makebox(0,0)[l]{\strut{} 55000}}}%
      \colorrgb{0.50,0.50,0.50}%
      \put(3731,100){\rotatebox{30}{\makebox(0,0)[l]{\strut{} 56000}}}%
      \colorrgb{0.50,0.50,0.50}%
      \put(4241,100){\rotatebox{30}{\makebox(0,0)[l]{\strut{} 57000}}}%
      \csname LTb\endcsname%
      \put(80,1036){\rotatebox{-270}{\makebox(0,0){\strut{} Residual ($\mu$s)}}}%
    }%
    \gplgaddtomacro\gplfronttext{%
    }%
    \gplbacktext
    \put(0,0){\includegraphics{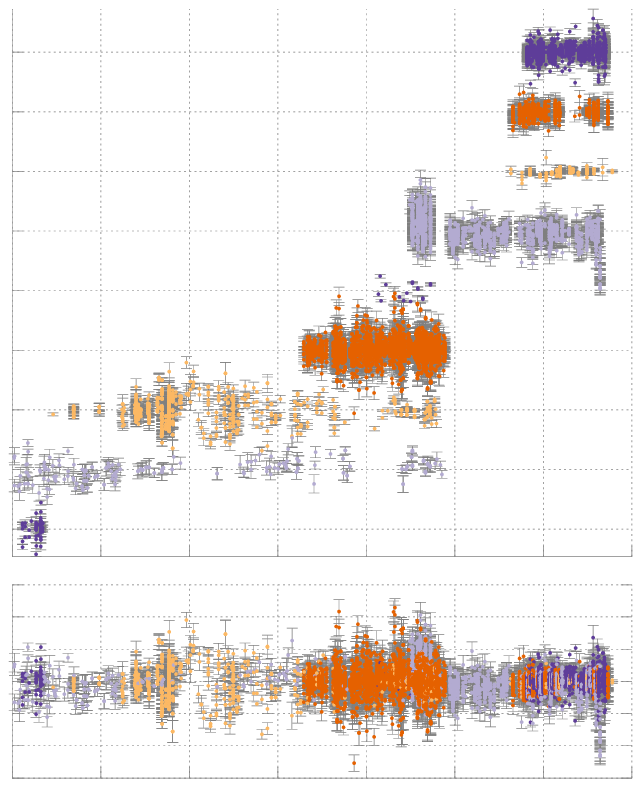}}%
    \gplfronttext
  \end{picture}%
\endgroup

%% file: tables/rcvr.tex
\begin{table}
\caption{Telescope and receiver-wise description  of the dataset, showing the bandwidth (BW), the centre frequency of observations ($f_c$), the number of ToAs retained after the the selection process described in the text and the MJD ranges over which the ToAs exist. For the older backends (see text), only ToAs were available. For the new backends archives were processed as described in \autoref{sec:obs_analysis}. \label{tab:rcvr}}
\begin{tabular}{l*{4}{l}}
\hline \hline
\multicolumn{5}{c}{\textsc {Telescopes and Receivers used for the dataset } }\\
\hline \hline
Telescope&\phantom{0}BW&\phantom{00}$f_c$&No. of&MJD\\
+Backend&(MHz)&(MHz)&ToAs&range\\
\hline
Effelsberg&\phantom{00}28&\phantom{00}840& \phantom{000}24& 51772--53159 \\
+EBPP&\phantom{00}56&\phantom{0}1410&\phantom{00}690&50460--54791\\
&\phantom{0}112&\phantom{0}2639&\phantom{000}35&51247--54690\\
\\
+PSRIX&\phantom{0}200&\phantom{0}1360&\phantom{00}120&55759--56494\\
&\phantom{0}100&\phantom{0}2640&\phantom{00}116&55632--56779 \\
\hline
Lovell&\phantom{00}16&\phantom{00}410&\phantom{0000}8& 50157--50695 \\
+A/DFB&\phantom{00}16&\phantom{00}610&\phantom{000}42& 50021--51452 \\
(see note below)&\phantom{00}16&\phantom{0}1400&\phantom{00}154& 49989--54853\\
\\
+ROACH&\phantom{0}400&\phantom{0}1532&\phantom{00}844& 55656--56729 \\
\hline
Nan\c{c}ay+BON&\phantom{0}128&\phantom{0}1397&\phantom{0}4502& 53293--54888 \\
\\
+NUPPI&\phantom{0}512&\phantom{0}1484&\phantom{0}2324& 55817--56700 \\
\hline
WSRT+PuMa&\phantom{00}80&\phantom{0}1380&\phantom{000}20& 54135--54372 \\
\\
+PuMaII&\phantom{00}80&\phantom{00}345&\phantom{0}1173& 54490--56640 \\
&\phantom{0}160&\phantom{0}1380&\phantom{00}536& 54520--56640 \\
\hline
Parkes&\phantom{0}128&\phantom{0}1400&\phantom{000}23& 50116--50343 \\
+FPTM&\phantom{0}128&\phantom{0}1700&\phantom{000}31&50116--50343\\
\hline \hline          
\end{tabular}
Note : The figures for bandwith (BW) and centre frequency ($f_c$) for the Jodrell Bank A/DFB and Parkes data are indicative only since the observations were made with various configurations. Details for these can be found in \cite{sbm+98}. Similar details for the other telescopes can be found in \cite{DCL+16}, \cite{BJK+16} or other specific references listed in \autoref{sec:obs_analysis}. 
\end{table}

%% file: tables/aeonwise.tex
\begin{table}
\caption{Properties of the TOA sets for each individual aeon ($\sim$1095 MJD period), determined using the respective ELL1 models. Note: The reduced $\chi^{2}$ values shown below are derived after applying error scaling or EFACs as described in \protect \autoref{sec:obs_analysis} \label{tab:aeonwise}}
\begin{tabular}{c*{3}{c}}
\hline \hline
\multicolumn{4}{c}{\textsc {Aeon-wise properties of dataset } }\\
\hline \hline
MJD Range  & {\small Weighted RMS} & Reduced & Number \\
&{\small Timing Residual} & $\chi ^2$ value &  of ToAs \\
&($\mu$s)&  &  \\
\hline
49989 -- 51062 & \phantom{0}8.9& 1.0 &\phantom{0}143 \\
50724 -- 51812 & 13.2& 1.0 &\phantom{0}331 \\
51451 -- 52538 & 14.2& 0.9 &\phantom{0}195 \\
52213 -- 53258 & 19.3& 1.0 &\phantom{0}146 \\
52927 -- 54004 & \phantom{0}9.2& 1.5 &1037 \\
53643 -- 54733 & \phantom{0}9.5& 1.0 &2518 \\
54372 -- 55444 & 10.8& 1.4 &1959 \\
55121 -- 56189 & \phantom{0}5.0& 1.1 &1679 \\
55836 -- 56880 & \phantom{0}6.2& 0.9 &2110 \\
\hline \hline
\end{tabular}
\end{table}

%% file: tables/compare.tex
\begin{table*}
\caption{Comparision of selected parameters of the black-widow pulsar system J2051$-$0827 with published values. $\mu_{\alpha}$ and $\mu_{\delta}$ values for the ELL1 model are obtained from a weighted fit to the position measured at succesive aeons. The epoch of DM determination need not correspond with the epoch of the timing model since the DM value for the ELL1 models are fixed from \protect \cite{kvh+16}, as explained in \autoref{ssec:dmvar}. Similarly, the DM value used in the \protect \cite{lvt+11} analysis is taken from \protect \cite{sbm+98}.\label{tab:compare}}
\begin{tabular}{lc*{3}{c}}
\hline \hline
Parameter  & \cite{dlk+01} & \cite{lvt+11} & ELL1 Model (Best fit) \\
\hline
MJD range of timing model fit& 49573 -- 51908& 53293 -- 54888& 55121.8 -- 56189.9\\
Proper motion in R.A., $\mu_{\alpha} \: {\rm (mas \: yr^{-1})}$ & $ 5.3(10)$ & 6.6(2) &  5.63(4) \\
Proper motion in Decl., $\mu_{\delta} \: {\rm (mas \: yr^{-1})}$ & $ 0.3(30)$ & 3.2(7) & 2.34(28) \\
Dispersion measure, DM (${\rm pc\: cm^{-3}}$) & $20.7449(4)$ & 20.7458(2) & 20.7299(17) \\
Epoch of DM measurement, (MJD) & 51000.0 & 49530.0 & 56387.8 \\
Eccentricity, $e$   &   $ < 9.6\times 10^{-5}$ & $6(1)\times 10^{-5}$ & $5.1(8) \times 10^{-5}$ \\
Reduced $\chi ^2$ (with scaled uncertainties)& -- & 1 & 1.1 \\
Number of TOAs  &  584  & 3126 & 1679 \\
Solar-system ephemeris model & DE200 & DE405 & DE421 \\
Timing Software Package& TIMAPR/\textsc{Tempo} & \textsc{Tempo2} & \textsc{Tempo2} \\
\hline \hline
\end{tabular}
\end{table*}

%% file: tables/merged.tex
\begin{table}	
\caption{Timing parameters for \psr{} for the ELL1 (implemented via the \textsc{Tempo2} hybrid model T2) and the BTX models. The values of derived parameters are italicised while parameters that should be neccesarily excluded from the respective timing models are marked as N/A. Note that the DM values presented here are obtained from \protect \cite{kvh+16}. For brevity, the table below uses the following abbreviations: FB0 indicates orbital frequency and higher numbers the resp. derivative, NToA denotes the number of TOAs, RMS $t_{\rm{resid}}$ denotes the RMS timing residual and Red. $\chi^{2}$ is the reduced $\chi^{2}$ value for the weighted \textsc{Tempo2} fit. $\tau_{\rm{char}}$ is the characteristic age associated with the pulsar and $B_{\rm{surf}}$ is the estimated surface magnetic field strength. The TT(TAI) clock correction procedure and the DE421 Solar Sytem Ephemerides were used for both the models. The units are in TCB \protect \citep[See][for details]{hem06}. The figures in parentheses are the nominal 1$\sigma$ \textsc{Tempo2} uncertainties in the least-significant digits quoted. The coordinates refer to J2000.\label{tab:ephem}}	
\begin{tabular}{@{}lll}	
\hline\hline	
Pulsar name & \multicolumn{2}{c}{\psr{}}\\
Binary model & T2 & BTX \\
MJD range & 55121.8--56189.9 & 49989.9--56779.3 \\ 
\parbox[c]{2.0cm}{\raggedright NToA }& 1679 & 11391 \\
\parbox[c]{2.0cm}{\raggedright RMS $t_{\rm{resid}}$ ($\mu s$)} & 5.0 & 5.2 \\
\parbox[c]{2.0cm}{\raggedright Red. $\chi^2$} & 1.1 & 4.2 \\
\parbox[c]{2.0cm}{\raggedright R. A., $\alpha$ }& 20:51:07.519768(18) & 20:51:07.519763(8) \\ 
\parbox[c]{2.0cm}{\raggedright Dec., $\delta$ }& $-$08:27:37.7497(8) & $-$08:27:37.7505(4) \\ 
\parbox[c]{2.0cm}{\raggedright Ec. Long., $\lambda$ }& 312.83572688(8)&312.83572710(2)\\ 
\parbox[c]{2.0cm}{\raggedright Ec. Lat., $\beta$ }&\phantom{00}8.8463418(5)&\phantom{00}8.84634230(9)  \\ 
$\mu_{\alpha}$ (mas\,yr$^{-1}$) & 5.63(4) & 5.57(4) \\ 
$\mu_{\delta}$ (mas\,yr$^{-1}$) & 2.34(28) & 3.60(10) \\ 
$\mu_{\lambda}$ (mas\,yr$^{-1}$) & 7.2(3) &6.34(1)\\ 
$\mu_{\beta}$ (mas\,yr$^{-1}$) & 4.6(23) &1.9(1)\\ 
$\nu$ (s$^{-1}$) & 221.796283653017(5) & 221.7962836530492(10) \\ 
$\dot{\nu}$ (s$^{-2}$) & $-$6.264(3)$\times 10^{-16}$ & $-$6.26532(6)$\times 10^{-16}$ \\ 
$P$ (ms) & {\em 4.50864182000643(8)} & {\em 4.5086418200061(5)} \\
$\dot{P}$ & {\em 1.2732(4)$\times 10^{-20}$} & {\em 1.27374(3)$\times 10^{-20}$} \\ 	
DM (cm$^{-3}$pc) & 20.7299(17) & 20.7299(17) \\ 	
$x$ (lt-s) & 0.0450720(3) & 0.04507074(20) \\	 
$\dot{x}$ & 1.3(148)$\times 10^{-16} {\phantom{+}}$ & 9.6(12)$\times 10^{-15}$\\ 	
$P_b$ (d) & 0.09911025490(4) &  N/A \\ 	
$\dot{P_b}$ & -5.9(3)$\times 10^{-12}$ &  N/A\\ 	
FB0(Hz)\ &  N/A & 1.1677979406(7)$\times 10^{-4}$ \\ 
FB1$(s^{-2})$\ & N/A & 8.2(4)$\times 10^{-20}$ \\ 
FB2$(s^{-3})$\ & N/A & $-$7.4(3)$\times 10^{-27}$ \\ 
FB3$(s^{-4})$\ & N/A & $-$6.3(16)$\times 10^{-35}$ \\ 
FB4$(s^{-5})$\ & N/A & 3.9(8)$\times 10^{-42}$ \\ 
FB5$(s^{-6})$\ & N/A & 1.8(7)$\times 10^{-49}$ \\ 
FB6$(s^{-7})$\ & N/A & 6.5(24)$\times 10^{-57}$ \\ 
FB7$(s^{-8})$\ & N/A & $-$5.8(23)$\times 10^{-64}$ \\ 
FB8$(s^{-9})$\ & N/A & $-$4.0(8)$\times 10^{-71}$ \\ 
FB9$(s^{-10})$\ & N/A & 1.6(7)$\times 10^{-78}$ \\ 
FB10$(s^{-11})$\ & N/A & 1.4(3)$\times 10^{-85}$ \\ 
FB11$(s^{-12})$\ & N/A & $-$3.2(18)$\times 10^{-93}$ \\ 
FB12$(s^{-13})$\ & N/A & $-$3.7(8)$\times 10^{-100}$ \\ 
FB13$(s^{-14})$\ & N/A & 3.0(30)$\times 10^{-108}$\\
FB14$(s^{-15})$\ & N/A & 7.3(19)$\times 10^{-115}$\\
FB15$(s^{-16})$\ & N/A & 5.2(20)$\times 10^{-123}$\\
FB16$(s^{-17})$\ & N/A & $-$7.9(25)$\times 10^{-130}$\\
FB17$(s^{-18})$\ & N/A & $-$1.8(5)$\times 10^{-137}$\\ 
\parbox[c]{2.0cm}{\raggedright Ref. epoch (MJD) }& 55655 & 55655 \\
$\omega$ (deg) & {\em 36(10)} & 0\\	
EPS1 & 3.0(10)$\times 10^{-5}$ & N/A \\ 	
EPS2 & 4.1(9)$\times 10^{-5}$ & N/A \\ 	
$e$ & {\em 5.1(8)$\times 10^{-5}$} & 0\\	
TASC (MJD) & 54091.0343079(8) & {\em 54091.03434936(14)} \\ 	
T0 (MJD) & {\em 54091.044(2)}& 54091.03434936(14) \\	
$\log_{10} \tau_{\rm{char}}$ (yr) & {\em 9.75} & {\em 9.75} \\	
$\log_{10} B_{\rm{surf}}$ (G) & {\em 8.38} & {\em 8.38} \\	
\hline	
\hline
\end{tabular}
\end{table}

%% file: plots/epslatex/RA_DEC_1095.tex
\begingroup
  \makeatletter
  \providecommand\color[2][]{%
    \GenericError{(gnuplot) \space\space\space\@spaces}{%
      Package color not loaded in conjunction with
      terminal option `colourtext'%
    }{See the gnuplot documentation for explanation.%
    }{Either use 'blacktext' in gnuplot or load the package
      color.sty in LaTeX.}%
    \renewcommand\color[2][]{}%
  }%
  \providecommand\includegraphics[2][]{%
    \GenericError{(gnuplot) \space\space\space\@spaces}{%
      Package graphicx or graphics not loaded%
    }{See the gnuplot documentation for explanation.%
    }{The gnuplot epslatex terminal needs graphicx.sty or graphics.sty.}%
    \renewcommand\includegraphics[2][]{}%
  }%
  \providecommand\rotatebox[2]{#2}%
  \@ifundefined{ifGPcolor}{%
    \newif\ifGPcolor
    \GPcolorfalse
  }{}%
  \@ifundefined{ifGPblacktext}{%
    \newif\ifGPblacktext
    \GPblacktexttrue
  }{}%
  \let\gplgaddtomacro\g@addto@macro
  \gdef\gplbacktext{}%
  \gdef\gplfronttext{}%
  \makeatother
  \ifGPblacktext
    \def\colorrgb#1{}%
    \def\colorgray#1{}%
  \else
    \ifGPcolor
      \def\colorrgb#1{\color[rgb]{#1}}%
      \def\colorgray#1{\color[gray]{#1}}%
      \expandafter\def\csname LTw\endcsname{\color{white}}%
      \expandafter\def\csname LTb\endcsname{\color{black}}%
      \expandafter\def\csname LTa\endcsname{\color{black}}%
      \expandafter\def\csname LT0\endcsname{\color[rgb]{1,0,0}}%
      \expandafter\def\csname LT1\endcsname{\color[rgb]{0,1,0}}%
      \expandafter\def\csname LT2\endcsname{\color[rgb]{0,0,1}}%
      \expandafter\def\csname LT3\endcsname{\color[rgb]{1,0,1}}%
      \expandafter\def\csname LT4\endcsname{\color[rgb]{0,1,1}}%
      \expandafter\def\csname LT5\endcsname{\color[rgb]{1,1,0}}%
      \expandafter\def\csname LT6\endcsname{\color[rgb]{0,0,0}}%
      \expandafter\def\csname LT7\endcsname{\color[rgb]{1,0.3,0}}%
      \expandafter\def\csname LT8\endcsname{\color[rgb]{0.5,0.5,0.5}}%
    \else
      \def\colorrgb#1{\color{black}}%
      \def\colorgray#1{\color[gray]{#1}}%
      \expandafter\def\csname LTw\endcsname{\color{white}}%
      \expandafter\def\csname LTb\endcsname{\color{black}}%
      \expandafter\def\csname LTa\endcsname{\color{black}}%
      \expandafter\def\csname LT0\endcsname{\color{black}}%
      \expandafter\def\csname LT1\endcsname{\color{black}}%
      \expandafter\def\csname LT2\endcsname{\color{black}}%
      \expandafter\def\csname LT3\endcsname{\color{black}}%
      \expandafter\def\csname LT4\endcsname{\color{black}}%
      \expandafter\def\csname LT5\endcsname{\color{black}}%
      \expandafter\def\csname LT6\endcsname{\color{black}}%
      \expandafter\def\csname LT7\endcsname{\color{black}}%
      \expandafter\def\csname LT8\endcsname{\color{black}}%
    \fi
  \fi
  \setlength{\unitlength}{0.0500bp}%
  \begin{picture}(4800.00,3600.00)%
    \gplgaddtomacro\gplbacktext{%
      \colorrgb{0.50,0.50,0.50}%
      \put(384,648){\makebox(0,0)[r]{\strut{}-5}}%
      \colorrgb{0.50,0.50,0.50}%
      \put(384,1059){\makebox(0,0)[r]{\strut{}-4}}%
      \colorrgb{0.50,0.50,0.50}%
      \put(384,1471){\makebox(0,0)[r]{\strut{}-3}}%
      \colorrgb{0.50,0.50,0.50}%
      \put(384,1882){\makebox(0,0)[r]{\strut{}-2}}%
      \colorrgb{0.50,0.50,0.50}%
      \put(384,2293){\makebox(0,0)[r]{\strut{}-1}}%
      \colorrgb{0.50,0.50,0.50}%
      \put(384,2704){\makebox(0,0)[r]{\strut{} 0}}%
      \colorrgb{0.50,0.50,0.50}%
      \put(384,3116){\makebox(0,0)[r]{\strut{} 1}}%
      \colorrgb{0.50,0.50,0.50}%
      \put(384,3527){\makebox(0,0)[r]{\strut{} 2}}%
      \colorrgb{0.50,0.50,0.50}%
      \put(192,300){\rotatebox{30}{\makebox(0,0)[l]{\strut{} 50000}}}%
      \colorrgb{0.50,0.50,0.50}%
      \put(452,300){\rotatebox{30}{\makebox(0,0)[l]{\strut{} 51000}}}%
      \colorrgb{0.50,0.50,0.50}%
      \put(712,300){\rotatebox{30}{\makebox(0,0)[l]{\strut{} 52000}}}%
      \colorrgb{0.50,0.50,0.50}%
      \put(972,300){\rotatebox{30}{\makebox(0,0)[l]{\strut{} 53000}}}%
      \colorrgb{0.50,0.50,0.50}%
      \put(1231,300){\rotatebox{30}{\makebox(0,0)[l]{\strut{} 54000}}}%
      \colorrgb{0.50,0.50,0.50}%
      \put(1491,300){\rotatebox{30}{\makebox(0,0)[l]{\strut{} 55000}}}%
      \colorrgb{0.50,0.50,0.50}%
      \put(1751,300){\rotatebox{30}{\makebox(0,0)[l]{\strut{} 56000}}}%
      \colorrgb{0.50,0.50,0.50}%
      \put(2011,300){\rotatebox{30}{\makebox(0,0)[l]{\strut{} 57000}}}%
      \csname LTb\endcsname%
      \put(84,2087){\rotatebox{-270}{\makebox(0,0){\strut{}$\Delta RA\ (\times 10^{-7}\ rad)$}}}%
    }%
    \gplgaddtomacro\gplfronttext{%
    }%
    \gplgaddtomacro\gplbacktext{%
      \colorrgb{0.50,0.50,0.50}%
      \put(2736,648){\makebox(0,0)[r]{\strut{}-3}}%
      \colorrgb{0.50,0.50,0.50}%
      \put(2736,1008){\makebox(0,0)[r]{\strut{}-2.5}}%
      \colorrgb{0.50,0.50,0.50}%
      \put(2736,1368){\makebox(0,0)[r]{\strut{}-2}}%
      \colorrgb{0.50,0.50,0.50}%
      \put(2736,1728){\makebox(0,0)[r]{\strut{}-1.5}}%
      \colorrgb{0.50,0.50,0.50}%
      \put(2736,2088){\makebox(0,0)[r]{\strut{}-1}}%
      \colorrgb{0.50,0.50,0.50}%
      \put(2736,2447){\makebox(0,0)[r]{\strut{}-0.5}}%
      \colorrgb{0.50,0.50,0.50}%
      \put(2736,2807){\makebox(0,0)[r]{\strut{} 0}}%
      \colorrgb{0.50,0.50,0.50}%
      \put(2736,3167){\makebox(0,0)[r]{\strut{} 0.5}}%
      \colorrgb{0.50,0.50,0.50}%
      \put(2736,3527){\makebox(0,0)[r]{\strut{} 1}}%
      \colorrgb{0.50,0.50,0.50}%
      \put(2544,300){\rotatebox{30}{\makebox(0,0)[l]{\strut{} 50000}}}%
      \colorrgb{0.50,0.50,0.50}%
      \put(2804,300){\rotatebox{30}{\makebox(0,0)[l]{\strut{} 51000}}}%
      \colorrgb{0.50,0.50,0.50}%
      \put(3064,300){\rotatebox{30}{\makebox(0,0)[l]{\strut{} 52000}}}%
      \colorrgb{0.50,0.50,0.50}%
      \put(3324,300){\rotatebox{30}{\makebox(0,0)[l]{\strut{} 53000}}}%
      \colorrgb{0.50,0.50,0.50}%
      \put(3583,300){\rotatebox{30}{\makebox(0,0)[l]{\strut{} 54000}}}%
      \colorrgb{0.50,0.50,0.50}%
      \put(3843,300){\rotatebox{30}{\makebox(0,0)[l]{\strut{} 55000}}}%
      \colorrgb{0.50,0.50,0.50}%
      \put(4103,300){\rotatebox{30}{\makebox(0,0)[l]{\strut{} 56000}}}%
      \colorrgb{0.50,0.50,0.50}%
      \put(4363,300){\rotatebox{30}{\makebox(0,0)[l]{\strut{} 57000}}}%
      \csname LTb\endcsname%
      \put(2400,2087){\rotatebox{-270}{\makebox(0,0){\strut{}$\Delta DEC\ (\times 10^{-7}\ rad)$}}}%
    }%
    \gplgaddtomacro\gplfronttext{%
    }%
    \gplbacktext
    \put(0,0){\includegraphics{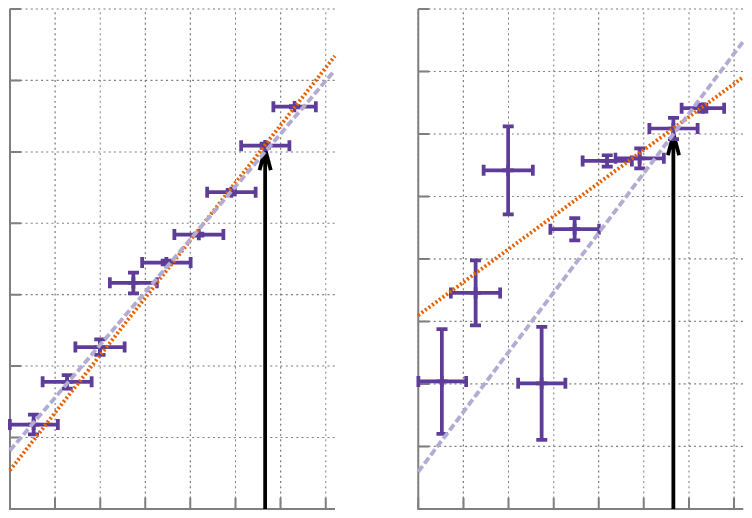}}%
    \gplfronttext
  \end{picture}%
\endgroup

%% file: plots/epslatex/dmvar.tex
\begingroup
  \makeatletter
  \providecommand\color[2][]{%
    \GenericError{(gnuplot) \space\space\space\@spaces}{%
      Package color not loaded in conjunction with
      terminal option `colourtext'%
    }{See the gnuplot documentation for explanation.%
    }{Either use 'blacktext' in gnuplot or load the package
      color.sty in LaTeX.}%
    \renewcommand\color[2][]{}%
  }%
  \providecommand\includegraphics[2][]{%
    \GenericError{(gnuplot) \space\space\space\@spaces}{%
      Package graphicx or graphics not loaded%
    }{See the gnuplot documentation for explanation.%
    }{The gnuplot epslatex terminal needs graphicx.sty or graphics.sty.}%
    \renewcommand\includegraphics[2][]{}%
  }%
  \providecommand\rotatebox[2]{#2}%
  \@ifundefined{ifGPcolor}{%
    \newif\ifGPcolor
    \GPcolortrue
  }{}%
  \@ifundefined{ifGPblacktext}{%
    \newif\ifGPblacktext
    \GPblacktexttrue
  }{}%
  \let\gplgaddtomacro\g@addto@macro
  \gdef\gplbacktext{}%
  \gdef\gplfronttext{}%
  \makeatother
  \ifGPblacktext
    \def\colorrgb#1{}%
    \def\colorgray#1{}%
  \else
    \ifGPcolor
      \def\colorrgb#1{\color[rgb]{#1}}%
      \def\colorgray#1{\color[gray]{#1}}%
      \expandafter\def\csname LTw\endcsname{\color{white}}%
      \expandafter\def\csname LTb\endcsname{\color{black}}%
      \expandafter\def\csname LTa\endcsname{\color{black}}%
      \expandafter\def\csname LT0\endcsname{\color[rgb]{1,0,0}}%
      \expandafter\def\csname LT1\endcsname{\color[rgb]{0,1,0}}%
      \expandafter\def\csname LT2\endcsname{\color[rgb]{0,0,1}}%
      \expandafter\def\csname LT3\endcsname{\color[rgb]{1,0,1}}%
      \expandafter\def\csname LT4\endcsname{\color[rgb]{0,1,1}}%
      \expandafter\def\csname LT5\endcsname{\color[rgb]{1,1,0}}%
      \expandafter\def\csname LT6\endcsname{\color[rgb]{0,0,0}}%
      \expandafter\def\csname LT7\endcsname{\color[rgb]{1,0.3,0}}%
      \expandafter\def\csname LT8\endcsname{\color[rgb]{0.5,0.5,0.5}}%
    \else
      \def\colorrgb#1{\color{black}}%
      \def\colorgray#1{\color[gray]{#1}}%
      \expandafter\def\csname LTw\endcsname{\color{white}}%
      \expandafter\def\csname LTb\endcsname{\color{black}}%
      \expandafter\def\csname LTa\endcsname{\color{black}}%
      \expandafter\def\csname LT0\endcsname{\color{black}}%
      \expandafter\def\csname LT1\endcsname{\color{black}}%
      \expandafter\def\csname LT2\endcsname{\color{black}}%
      \expandafter\def\csname LT3\endcsname{\color{black}}%
      \expandafter\def\csname LT4\endcsname{\color{black}}%
      \expandafter\def\csname LT5\endcsname{\color{black}}%
      \expandafter\def\csname LT6\endcsname{\color{black}}%
      \expandafter\def\csname LT7\endcsname{\color{black}}%
      \expandafter\def\csname LT8\endcsname{\color{black}}%
    \fi
  \fi
  \setlength{\unitlength}{0.0500bp}%
  \begin{picture}(4818.00,3400.00)%
    \gplgaddtomacro\gplbacktext{%
      \colorrgb{0.50,0.50,0.50}%
      \put(385,408){\makebox(0,0)[r]{\strut{} -2.0}}%
      \colorrgb{0.50,0.50,0.50}%
      \put(385,826){\makebox(0,0)[r]{\strut{} -1.5}}%
      \colorrgb{0.50,0.50,0.50}%
      \put(385,1243){\makebox(0,0)[r]{\strut{} -1.0}}%
      \colorrgb{0.50,0.50,0.50}%
      \put(385,1661){\makebox(0,0)[r]{\strut{} -0.5}}%
      \colorrgb{0.50,0.50,0.50}%
      \put(385,2078){\makebox(0,0)[r]{\strut{} 0}}%
      \colorrgb{0.50,0.50,0.50}%
      \put(385,2496){\makebox(0,0)[r]{\strut{} 0.5}}%
      \colorrgb{0.50,0.50,0.50}%
      \put(385,2913){\makebox(0,0)[r]{\strut{} 1.0}}%
      \colorrgb{0.50,0.50,0.50}%
      \put(385,3331){\makebox(0,0)[r]{\strut{} 1.5}}%
      \colorrgb{0.50,0.50,0.50}%
      \put(193,100){\rotatebox{30}{\makebox(0,0)[l]{\strut{} 54600}}}%
      \colorrgb{0.50,0.50,0.50}%
      \put(578,100){\rotatebox{30}{\makebox(0,0)[l]{\strut{} 54800}}}%
      \colorrgb{0.50,0.50,0.50}%
      \put(964,100){\rotatebox{30}{\makebox(0,0)[l]{\strut{} 55000}}}%
      \colorrgb{0.50,0.50,0.50}%
      \put(1349,100){\rotatebox{30}{\makebox(0,0)[l]{\strut{} 55200}}}%
      \colorrgb{0.50,0.50,0.50}%
      \put(1734,100){\rotatebox{30}{\makebox(0,0)[l]{\strut{} 55400}}}%
      \colorrgb{0.50,0.50,0.50}%
      \put(2120,100){\rotatebox{30}{\makebox(0,0)[l]{\strut{} 55600}}}%
      \colorrgb{0.50,0.50,0.50}%
      \put(2505,100){\rotatebox{30}{\makebox(0,0)[l]{\strut{} 55800}}}%
      \colorrgb{0.50,0.50,0.50}%
      \put(2891,100){\rotatebox{30}{\makebox(0,0)[l]{\strut{} 56000}}}%
      \colorrgb{0.50,0.50,0.50}%
      \put(3276,100){\rotatebox{30}{\makebox(0,0)[l]{\strut{} 56200}}}%
      \colorrgb{0.50,0.50,0.50}%
      \put(3661,100){\rotatebox{30}{\makebox(0,0)[l]{\strut{} 56400}}}%
      \colorrgb{0.50,0.50,0.50}%
      \put(4047,100){\rotatebox{30}{\makebox(0,0)[l]{\strut{} 56600}}}%
      \colorrgb{0.50,0.50,0.50}%
      \put(4432,100){\rotatebox{30}{\makebox(0,0)[l]{\strut{} 56800}}}%
      \csname LTb\endcsname%
      \put(50,2069){\rotatebox{-270}{\makebox(0,0){\strut{}$\Delta DM(\times 10^{-3}cm^{-3}pc)$}}}%
      \put(2505,0){\rotatebox{0}{\makebox(0,0){\strut{}MJD}}}%
    }%
    \gplgaddtomacro\gplfronttext{%
    }%
    \gplbacktext
    \put(0,0){\includegraphics{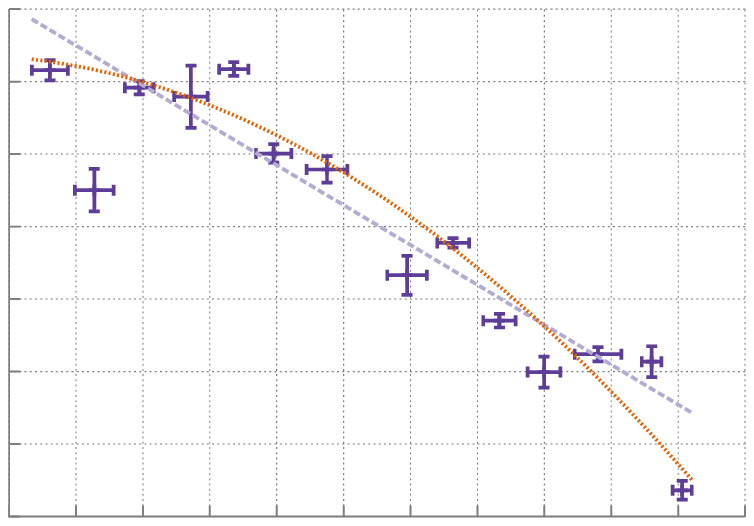}}%
    \gplfronttext
  \end{picture}%
\endgroup

%% file: plots/epslatex/year.tex
\begingroup
  \makeatletter
  \providecommand\color[2][]{%
    \GenericError{(gnuplot) \space\space\space\@spaces}{%
      Package color not loaded in conjunction with
      terminal option `colourtext'%
    }{See the gnuplot documentation for explanation.%
    }{Either use 'blacktext' in gnuplot or load the package
      color.sty in LaTeX.}%
    \renewcommand\color[2][]{}%
  }%
  \providecommand\includegraphics[2][]{%
    \GenericError{(gnuplot) \space\space\space\@spaces}{%
      Package graphicx or graphics not loaded%
    }{See the gnuplot documentation for explanation.%
    }{The gnuplot epslatex terminal needs graphicx.sty or graphics.sty.}%
    \renewcommand\includegraphics[2][]{}%
  }%
  \providecommand\rotatebox[2]{#2}%
  \@ifundefined{ifGPcolor}{%
    \newif\ifGPcolor
    \GPcolortrue
  }{}%
  \@ifundefined{ifGPblacktext}{%
    \newif\ifGPblacktext
    \GPblacktexttrue
  }{}%
  \let\gplgaddtomacro\g@addto@macro
  \gdef\gplbacktext{}%
  \gdef\gplfronttext{}%
  \makeatother
  \ifGPblacktext
    \def\colorrgb#1{}%
    \def\colorgray#1{}%
  \else
    \ifGPcolor
      \def\colorrgb#1{\color[rgb]{#1}}%
      \def\colorgray#1{\color[gray]{#1}}%
      \expandafter\def\csname LTw\endcsname{\color{white}}%
      \expandafter\def\csname LTb\endcsname{\color{black}}%
      \expandafter\def\csname LTa\endcsname{\color{black}}%
      \expandafter\def\csname LT0\endcsname{\color[rgb]{1,0,0}}%
      \expandafter\def\csname LT1\endcsname{\color[rgb]{0,1,0}}%
      \expandafter\def\csname LT2\endcsname{\color[rgb]{0,0,1}}%
      \expandafter\def\csname LT3\endcsname{\color[rgb]{1,0,1}}%
      \expandafter\def\csname LT4\endcsname{\color[rgb]{0,1,1}}%
      \expandafter\def\csname LT5\endcsname{\color[rgb]{1,1,0}}%
      \expandafter\def\csname LT6\endcsname{\color[rgb]{0,0,0}}%
      \expandafter\def\csname LT7\endcsname{\color[rgb]{1,0.3,0}}%
      \expandafter\def\csname LT8\endcsname{\color[rgb]{0.5,0.5,0.5}}%
    \else
      \def\colorrgb#1{\color{black}}%
      \def\colorgray#1{\color[gray]{#1}}%
      \expandafter\def\csname LTw\endcsname{\color{white}}%
      \expandafter\def\csname LTb\endcsname{\color{black}}%
      \expandafter\def\csname LTa\endcsname{\color{black}}%
      \expandafter\def\csname LT0\endcsname{\color{black}}%
      \expandafter\def\csname LT1\endcsname{\color{black}}%
      \expandafter\def\csname LT2\endcsname{\color{black}}%
      \expandafter\def\csname LT3\endcsname{\color{black}}%
      \expandafter\def\csname LT4\endcsname{\color{black}}%
      \expandafter\def\csname LT5\endcsname{\color{black}}%
      \expandafter\def\csname LT6\endcsname{\color{black}}%
      \expandafter\def\csname LT7\endcsname{\color{black}}%
      \expandafter\def\csname LT8\endcsname{\color{black}}%
    \fi
  \fi
  \setlength{\unitlength}{0.0500bp}%
  \begin{picture}(4818.00,5668.00)%
    \gplgaddtomacro\gplbacktext{%
      \colorrgb{0.50,0.50,0.50}%
      \put(385,3174){\makebox(0,0)[r]{\strut{}-1}}%
      \colorrgb{0.50,0.50,0.50}%
      \put(385,3471){\makebox(0,0)[r]{\strut{}-0.5}}%
      \colorrgb{0.50,0.50,0.50}%
      \put(385,3769){\makebox(0,0)[r]{\strut{} 0}}%
      \colorrgb{0.50,0.50,0.50}%
      \put(385,4066){\makebox(0,0)[r]{\strut{} 0.5}}%
      \colorrgb{0.50,0.50,0.50}%
      \put(385,4364){\makebox(0,0)[r]{\strut{} 1}}%
      \colorrgb{0.50,0.50,0.50}%
      \put(385,4661){\makebox(0,0)[r]{\strut{} 1.5}}%
      \colorrgb{0.50,0.50,0.50}%
      \put(385,4958){\makebox(0,0)[r]{\strut{} 2}}%
      \colorrgb{0.50,0.50,0.50}%
      \put(385,5256){\makebox(0,0)[r]{\strut{} 2.5}}%
      \colorrgb{0.50,0.50,0.50}%
      \put(385,5553){\makebox(0,0)[r]{\strut{} 3}}%
      \csname LTb\endcsname%
      \put(100,4363){\rotatebox{-270}{\makebox(0,0){\strut{}$\Delta P_{\rm b} (\times 10^{-8}\ days)$}}}%
    }%
    \gplgaddtomacro\gplfronttext{%
    }%
    \gplgaddtomacro\gplbacktext{%
      \colorrgb{0.50,0.50,0.50}%
      \put(385,680){\makebox(0,0)[r]{\strut{}-6}}%
      \colorrgb{0.50,0.50,0.50}%
      \put(385,1077){\makebox(0,0)[r]{\strut{}-4}}%
      \colorrgb{0.50,0.50,0.50}%
      \put(385,1473){\makebox(0,0)[r]{\strut{}-2}}%
      \colorrgb{0.50,0.50,0.50}%
      \put(385,1870){\makebox(0,0)[r]{\strut{} 0}}%
      \colorrgb{0.50,0.50,0.50}%
      \put(385,2267){\makebox(0,0)[r]{\strut{} 2}}%
      \colorrgb{0.50,0.50,0.50}%
      \put(385,2663){\makebox(0,0)[r]{\strut{} 4}}%
      \colorrgb{0.50,0.50,0.50}%
      \put(385,3060){\makebox(0,0)[r]{\strut{} 6}}%
      \colorrgb{0.50,0.50,0.50}%
      \put(193,350){\rotatebox{30}{\makebox(0,0)[l]{\strut{} 50000}}}%
      \colorrgb{0.50,0.50,0.50}%
      \put(799,350){\rotatebox{30}{\makebox(0,0)[l]{\strut{} 51000}}}%
      \colorrgb{0.50,0.50,0.50}%
      \put(1404,350){\rotatebox{30}{\makebox(0,0)[l]{\strut{} 52000}}}%
      \colorrgb{0.50,0.50,0.50}%
      \put(2010,350){\rotatebox{30}{\makebox(0,0)[l]{\strut{} 53000}}}%
      \colorrgb{0.50,0.50,0.50}%
      \put(2615,350){\rotatebox{30}{\makebox(0,0)[l]{\strut{} 54000}}}%
      \colorrgb{0.50,0.50,0.50}%
      \put(3221,350){\rotatebox{30}{\makebox(0,0)[l]{\strut{} 55000}}}%
      \colorrgb{0.50,0.50,0.50}%
      \put(3826,350){\rotatebox{30}{\makebox(0,0)[l]{\strut{} 56000}}}%
      \colorrgb{0.50,0.50,0.50}%
      \put(4432,350){\rotatebox{30}{\makebox(0,0)[l]{\strut{} 57000}}}%
      \csname LTb\endcsname%
      \put(100,1870){\rotatebox{-270}{\makebox(0,0){\strut{}$\Delta x (\times 10^{-5}\ lt \text{-} s)$}}}%
      \put(2615,200){\rotatebox{0}{\makebox(0,0){\strut{}MJD}}}%
    }%
    \gplgaddtomacro\gplfronttext{%
    }%
    \gplbacktext
    \put(0,0){\includegraphics{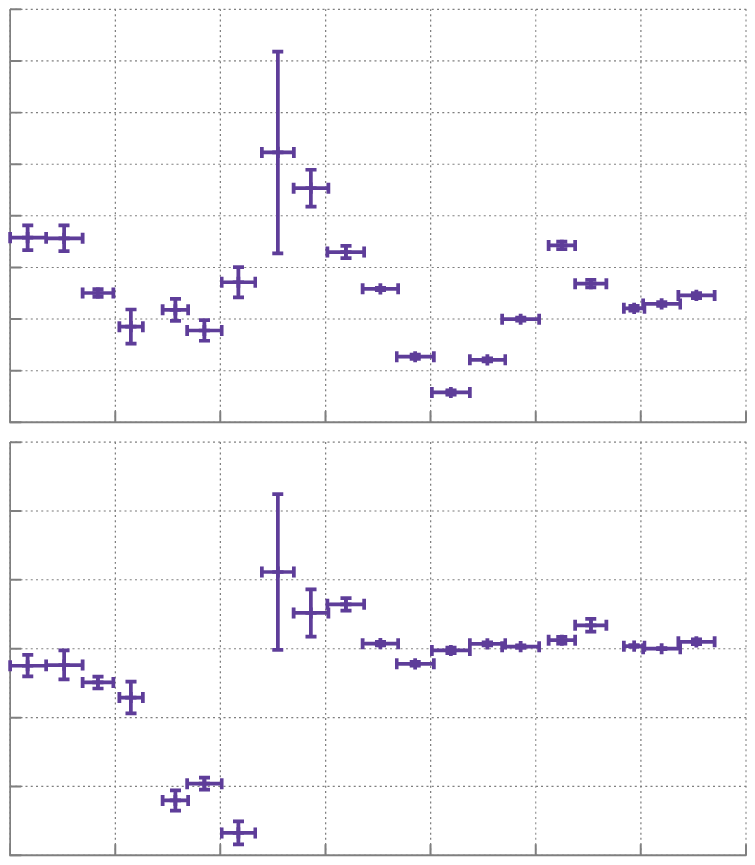}}%
    \gplfronttext
  \end{picture}%
\endgroup

%% file: plots/epslatex/tascyear.tex
\begingroup
  \makeatletter
  \providecommand\color[2][]{%
    \GenericError{(gnuplot) \space\space\space\@spaces}{%
      Package color not loaded in conjunction with
      terminal option `colourtext'%
    }{See the gnuplot documentation for explanation.%
    }{Either use 'blacktext' in gnuplot or load the package
      color.sty in LaTeX.}%
    \renewcommand\color[2][]{}%
  }%
  \providecommand\includegraphics[2][]{%
    \GenericError{(gnuplot) \space\space\space\@spaces}{%
      Package graphicx or graphics not loaded%
    }{See the gnuplot documentation for explanation.%
    }{The gnuplot epslatex terminal needs graphicx.sty or graphics.sty.}%
    \renewcommand\includegraphics[2][]{}%
  }%
  \providecommand\rotatebox[2]{#2}%
  \@ifundefined{ifGPcolor}{%
    \newif\ifGPcolor
    \GPcolortrue
  }{}%
  \@ifundefined{ifGPblacktext}{%
    \newif\ifGPblacktext
    \GPblacktexttrue
  }{}%
  \let\gplgaddtomacro\g@addto@macro
  \gdef\gplbacktext{}%
  \gdef\gplfronttext{}%
  \makeatother
  \ifGPblacktext
    \def\colorrgb#1{}%
    \def\colorgray#1{}%
  \else
    \ifGPcolor
      \def\colorrgb#1{\color[rgb]{#1}}%
      \def\colorgray#1{\color[gray]{#1}}%
      \expandafter\def\csname LTw\endcsname{\color{white}}%
      \expandafter\def\csname LTb\endcsname{\color{black}}%
      \expandafter\def\csname LTa\endcsname{\color{black}}%
      \expandafter\def\csname LT0\endcsname{\color[rgb]{1,0,0}}%
      \expandafter\def\csname LT1\endcsname{\color[rgb]{0,1,0}}%
      \expandafter\def\csname LT2\endcsname{\color[rgb]{0,0,1}}%
      \expandafter\def\csname LT3\endcsname{\color[rgb]{1,0,1}}%
      \expandafter\def\csname LT4\endcsname{\color[rgb]{0,1,1}}%
      \expandafter\def\csname LT5\endcsname{\color[rgb]{1,1,0}}%
      \expandafter\def\csname LT6\endcsname{\color[rgb]{0,0,0}}%
      \expandafter\def\csname LT7\endcsname{\color[rgb]{1,0.3,0}}%
      \expandafter\def\csname LT8\endcsname{\color[rgb]{0.5,0.5,0.5}}%
    \else
      \def\colorrgb#1{\color{black}}%
      \def\colorgray#1{\color[gray]{#1}}%
      \expandafter\def\csname LTw\endcsname{\color{white}}%
      \expandafter\def\csname LTb\endcsname{\color{black}}%
      \expandafter\def\csname LTa\endcsname{\color{black}}%
      \expandafter\def\csname LT0\endcsname{\color{black}}%
      \expandafter\def\csname LT1\endcsname{\color{black}}%
      \expandafter\def\csname LT2\endcsname{\color{black}}%
      \expandafter\def\csname LT3\endcsname{\color{black}}%
      \expandafter\def\csname LT4\endcsname{\color{black}}%
      \expandafter\def\csname LT5\endcsname{\color{black}}%
      \expandafter\def\csname LT6\endcsname{\color{black}}%
      \expandafter\def\csname LT7\endcsname{\color{black}}%
      \expandafter\def\csname LT8\endcsname{\color{black}}%
    \fi
  \fi
  \setlength{\unitlength}{0.0500bp}%
  \begin{picture}(10204.00,10204.00)%
    \gplgaddtomacro\gplbacktext{%
      \colorrgb{0.50,0.50,0.50}%
      \put(624,7442){\makebox(0,0)[r]{\strut{}-4}}%
      \colorrgb{0.50,0.50,0.50}%
      \put(624,7907){\makebox(0,0)[r]{\strut{}-2}}%
      \colorrgb{0.50,0.50,0.50}%
      \put(624,8372){\makebox(0,0)[r]{\strut{} 0}}%
      \colorrgb{0.50,0.50,0.50}%
      \put(624,8836){\makebox(0,0)[r]{\strut{} 2}}%
      \colorrgb{0.50,0.50,0.50}%
      \put(624,9301){\makebox(0,0)[r]{\strut{} 4}}%
      \colorrgb{0.50,0.50,0.50}%
      \put(624,9766){\makebox(0,0)[r]{\strut{} 6}}%
      \colorrgb{0.50,0.50,0.50}%
      \put(532,10100){\rotatebox{0}{\makebox(0,0)[l]{\strut{} Oct,1995}}}%
      \colorrgb{0.50,0.50,0.50}%
      \put(1872,10100){\rotatebox{0}{\makebox(0,0)[l]{\strut{} Jul,1998}}}%
      \colorrgb{0.50,0.50,0.50}%
      \put(3212,10100){\rotatebox{0}{\makebox(0,0)[l]{\strut{} Apr,2001}}}%
      \colorrgb{0.50,0.50,0.50}%
      \put(4552,10100){\rotatebox{0}{\makebox(0,0)[l]{\strut{} Dec,2003}}}%
      \colorrgb{0.50,0.50,0.50}%
      \put(5892,10100){\rotatebox{0}{\makebox(0,0)[l]{\strut{} Sep,2006}}}%
      \colorrgb{0.50,0.50,0.50}%
      \put(7232,10100){\rotatebox{0}{\makebox(0,0)[l]{\strut{} Jun,2009}}}%
      \colorrgb{0.50,0.50,0.50}%
      \put(8572,10100){\rotatebox{0}{\makebox(0,0)[l]{\strut{} Mar,2012}}}%
      %
      \csname LTb\endcsname%
      \put(224,8604){\rotatebox{-270}{\makebox(0,0){\strut{}$\Delta T_{\rm asc}\ (\times 10^{-5}\ days)$}}}%
    }%
    \gplgaddtomacro\gplfronttext{%
    }%
    \gplgaddtomacro\gplbacktext{%
      \colorrgb{0.50,0.50,0.50}%
      \put(624,4217){\makebox(0,0)[r]{\strut{}-2}}%
      \colorrgb{0.50,0.50,0.50}%
      \put(624,4527){\makebox(0,0)[r]{\strut{}-1.5}}%
      \colorrgb{0.50,0.50,0.50}%
      \put(624,4837){\makebox(0,0)[r]{\strut{}-1}}%
      \colorrgb{0.50,0.50,0.50}%
      \put(624,5147){\makebox(0,0)[r]{\strut{}-5}}%
      \colorrgb{0.50,0.50,0.50}%
      \put(624,5457){\makebox(0,0)[r]{\strut{} 0}}%
      \colorrgb{0.50,0.50,0.50}%
      \put(624,5766){\makebox(0,0)[r]{\strut{} 5}}%
      \colorrgb{0.50,0.50,0.50}%
      \put(624,6076){\makebox(0,0)[r]{\strut{} 1}}%
      \colorrgb{0.50,0.50,0.50}%
      \put(624,6386){\makebox(0,0)[r]{\strut{} 1.5}}%
      \colorrgb{0.50,0.50,0.50}%
      \put(624,6696){\makebox(0,0)[r]{\strut{} 2}}%
      \colorrgb{0.50,0.50,0.50}%
      \put(624,7006){\makebox(0,0)[r]{\strut{} 2.5}}%
      \csname LTb\endcsname%
      \put(224,5611){\rotatebox{-270}{\makebox(0,0){\strut{}$\Delta P_{\rm b}\ (\times 10^{-8}\ days)$}}}%
    }%
    \gplgaddtomacro\gplfronttext{%
    }%
    \gplgaddtomacro\gplbacktext{%
      \colorrgb{0.50,0.50,0.50}%
      \put(624,1490){\makebox(0,0)[r]{\strut{}-6}}%
      \colorrgb{0.50,0.50,0.50}%
      \put(624,2021){\makebox(0,0)[r]{\strut{}-4}}%
      \colorrgb{0.50,0.50,0.50}%
      \put(624,2552){\makebox(0,0)[r]{\strut{}-2}}%
      \colorrgb{0.50,0.50,0.50}%
      \put(624,3083){\makebox(0,0)[r]{\strut{} 0}}%
      \colorrgb{0.50,0.50,0.50}%
      \put(624,3615){\makebox(0,0)[r]{\strut{} 2}}%
      \colorrgb{0.50,0.50,0.50}%
      \put(432,968){\rotatebox{0}{\makebox(0,0)[l]{\strut{} 50000}}}%
      \colorrgb{0.50,0.50,0.50}%
      \put(1772,968){\rotatebox{0}{\makebox(0,0)[l]{\strut{} 51000}}}%
      \colorrgb{0.50,0.50,0.50}%
      \put(3112,968){\rotatebox{0}{\makebox(0,0)[l]{\strut{} 52000}}}%
      \colorrgb{0.50,0.50,0.50}%
      \put(4452,968){\rotatebox{0}{\makebox(0,0)[l]{\strut{} 53000}}}%
      \colorrgb{0.50,0.50,0.50}%
      \put(5792,968){\rotatebox{0}{\makebox(0,0)[l]{\strut{} 54000}}}%
      \colorrgb{0.50,0.50,0.50}%
      \put(7132,968){\rotatebox{0}{\makebox(0,0)[l]{\strut{} 55000}}}%
      \colorrgb{0.50,0.50,0.50}%
      \put(8472,968){\rotatebox{0}{\makebox(0,0)[l]{\strut{} 56000}}}%
      \csname LTb\endcsname%
      \put(224,2618){\rotatebox{-270}{\makebox(0,0){\strut{}$\Delta x\ (\times 10^{-5}\ lt\text{-}s)$}}}%
      \put(5300,600){\rotatebox{0}{\makebox(0,0){\strut{}MJD}}}%
    }%
    \gplgaddtomacro\gplfronttext{%
    }%
    \gplbacktext
    \put(-300,0){\includegraphics{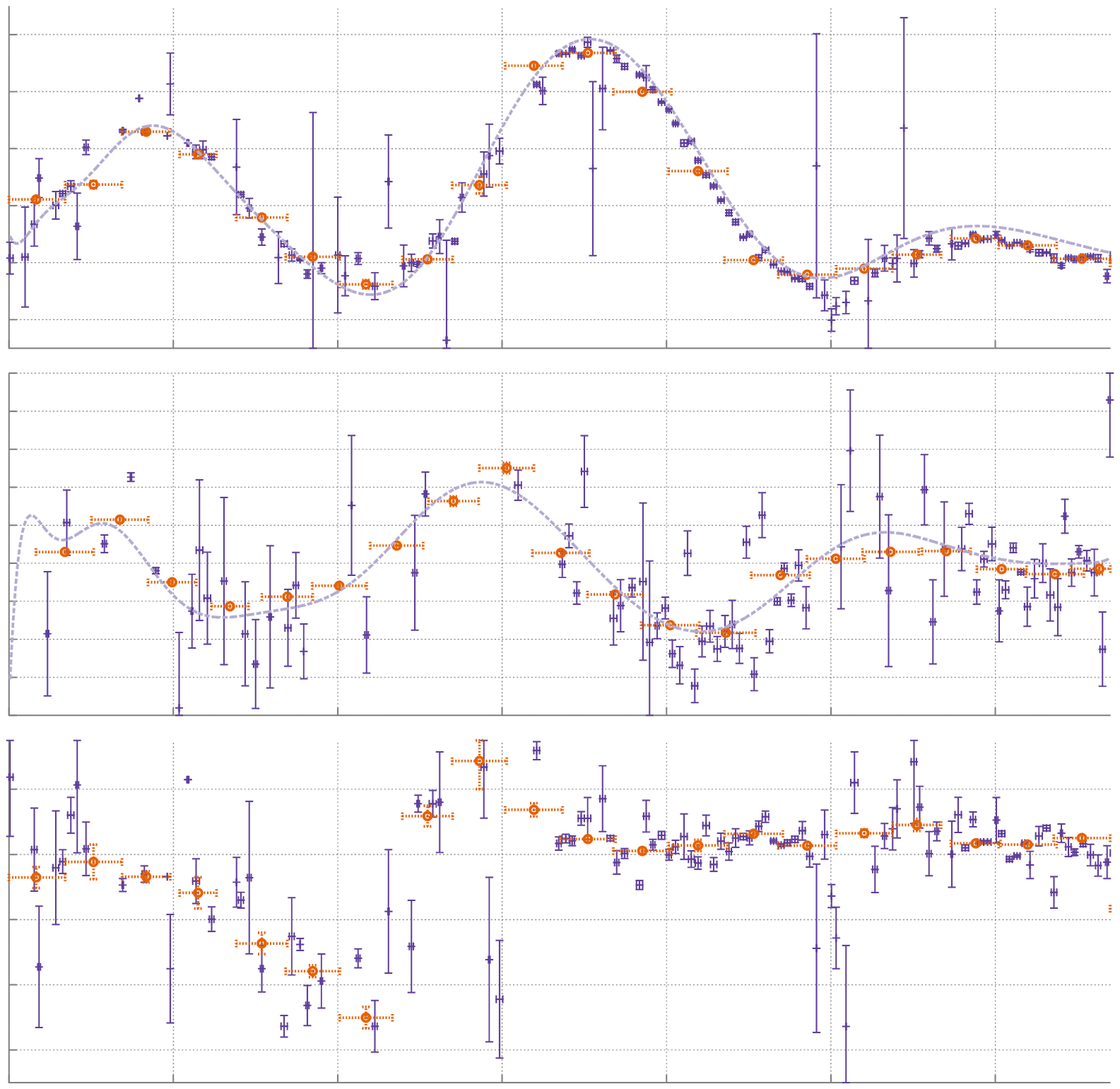}}%
    \gplfronttext
  \end{picture}%
\endgroup

%% file: tables/annuvar.tex
\begin{table}
\caption{Maximum contributions from the various sources of secular variations in $P_{\rm b}$ and $x$ as presented in \autoref{eq:Pb_terms} and \ref{eq:x_terms}. \label{tab:annuvar}}
\begin{tabular}{l*{2}{c}}
\hline\hline
Source &$\rm{\dot{P}_b}$ (days)&$\rm{\dot{x}}$ (lt\text{-}s)\\
\hline
GW emission&{$\PbGW{}$}&{$\xGW{}$}\\
Doppler Correction&{$\Pbdoppler{}$}&{$\xdoppler{}$}\\
Proper Motion Corr.& N/A &{\phantom{+}$\xPM{}$}\\
Varying Aberration& N/A	&{$\xdea{}$}\\
Mass loss&\multicolumn{2}{|c|}{Requires unphysical $\dot{m}_{\rm c} \sim 10^{-7}M_{\odot}$}\\
Mass/GQ Variations&\multicolumn{2}{|c|}{\xQ{}}\\
Spin-Orbit Coupling&\multicolumn{2}{|c|}{\xQ{}}\\
\hline
Max. Measured&$\phantom{+}\PBDmax{}$& $\phantom{+}\XDmax{}$\\
Min. Measured&$\PBDmin{}$&$\XDmin{}$\\
\hline\hline
\end{tabular}
Notes: \\
1. The contribution from qravitation quadrupole (GQ) and the classical spin-orbit coupling (SOC) variations require assumptions based on \cite{lvt+11}. Since the derived values are then identical to those presented there, readers are referred to the original source instead. \\
2. $\dot{m}_{\rm c}$ refers to the rate at which mass is lost by the companion.
\end{table}

%% file: plots/epslatex/tascyearzoom2.tex
\begingroup
  \makeatletter
  \providecommand\color[2][]{%
    \GenericError{(gnuplot) \space\space\space\@spaces}{%
      Package color not loaded in conjunction with
      terminal option `colourtext'%
    }{See the gnuplot documentation for explanation.%
    }{Either use 'blacktext' in gnuplot or load the package
      color.sty in LaTeX.}%
    \renewcommand\color[2][]{}%
  }%
  \providecommand\includegraphics[2][]{%
    \GenericError{(gnuplot) \space\space\space\@spaces}{%
      Package graphicx or graphics not loaded%
    }{See the gnuplot documentation for explanation.%
    }{The gnuplot epslatex terminal needs graphicx.sty or graphics.sty.}%
    \renewcommand\includegraphics[2][]{}%
  }%
  \providecommand\rotatebox[2]{#2}%
  \@ifundefined{ifGPcolor}{%
    \newif\ifGPcolor
    \GPcolortrue
  }{}%
  \@ifundefined{ifGPblacktext}{%
    \newif\ifGPblacktext
    \GPblacktexttrue
  }{}%
  \let\gplgaddtomacro\g@addto@macro
  \gdef\gplbacktext{}%
  \gdef\gplfronttext{}%
  \makeatother
  \ifGPblacktext
    \def\colorrgb#1{}%
    \def\colorgray#1{}%
  \else
    \ifGPcolor
      \def\colorrgb#1{\color[rgb]{#1}}%
      \def\colorgray#1{\color[gray]{#1}}%
      \expandafter\def\csname LTw\endcsname{\color{white}}%
      \expandafter\def\csname LTb\endcsname{\color{black}}%
      \expandafter\def\csname LTa\endcsname{\color{black}}%
      \expandafter\def\csname LT0\endcsname{\color[rgb]{1,0,0}}%
      \expandafter\def\csname LT1\endcsname{\color[rgb]{0,1,0}}%
      \expandafter\def\csname LT2\endcsname{\color[rgb]{0,0,1}}%
      \expandafter\def\csname LT3\endcsname{\color[rgb]{1,0,1}}%
      \expandafter\def\csname LT4\endcsname{\color[rgb]{0,1,1}}%
      \expandafter\def\csname LT5\endcsname{\color[rgb]{1,1,0}}%
      \expandafter\def\csname LT6\endcsname{\color[rgb]{0,0,0}}%
      \expandafter\def\csname LT7\endcsname{\color[rgb]{1,0.3,0}}%
      \expandafter\def\csname LT8\endcsname{\color[rgb]{0.5,0.5,0.5}}%
    \else
      \def\colorrgb#1{\color{black}}%
      \def\colorgray#1{\color[gray]{#1}}%
      \expandafter\def\csname LTw\endcsname{\color{white}}%
      \expandafter\def\csname LTb\endcsname{\color{black}}%
      \expandafter\def\csname LTa\endcsname{\color{black}}%
      \expandafter\def\csname LT0\endcsname{\color{black}}%
      \expandafter\def\csname LT1\endcsname{\color{black}}%
      \expandafter\def\csname LT2\endcsname{\color{black}}%
      \expandafter\def\csname LT3\endcsname{\color{black}}%
      \expandafter\def\csname LT4\endcsname{\color{black}}%
      \expandafter\def\csname LT5\endcsname{\color{black}}%
      \expandafter\def\csname LT6\endcsname{\color{black}}%
      \expandafter\def\csname LT7\endcsname{\color{black}}%
      \expandafter\def\csname LT8\endcsname{\color{black}}%
    \fi
  \fi
  \setlength{\unitlength}{0.0500bp}%
  \begin{picture}(4818.00,9070.00)%
    \gplgaddtomacro\gplbacktext{%
      \colorrgb{0.50,0.50,0.50}%
      \put(385,5396){\makebox(0,0)[r]{\strut{}-4}}%
      \colorrgb{0.50,0.50,0.50}%
      \put(385,6031){\makebox(0,0)[r]{\strut{}-2}}%
      \colorrgb{0.50,0.50,0.50}%
      \put(385,6666){\makebox(0,0)[r]{\strut{} 0}}%
      \colorrgb{0.50,0.50,0.50}%
      \put(385,7300){\makebox(0,0)[r]{\strut{} 2}}%
      \colorrgb{0.50,0.50,0.50}%
      \put(385,7935){\makebox(0,0)[r]{\strut{} 4}}%
      \colorrgb{0.50,0.50,0.50}%
      \put(385,8570){\makebox(0,0)[r]{\strut{} 6}}%
      \csname LTb\endcsname%
      \put(0,6983){\rotatebox{-270}{\makebox(0,0){\strut{}$\Delta T_{\rm asc} (\times 10^{-5}\ days)$}}}%
      \put(1130,8000){\rotatebox{-270}{\makebox(0,0){\strut{}{\large $\boldsymbol{\searrow}$}}}}%
      \put(1760,7483){\rotatebox{-270}{\makebox(0,0){\strut{}{\large $\boldsymbol{\nwarrow}$}}}}%
      \put(2180,6473){\rotatebox{-270}{\makebox(0,0){\strut{}{\large $\boldsymbol{\nwarrow}$}}}}%
      \put(2300,6280){\rotatebox{-270}{\makebox(0,0){\strut{}{\large $\boldsymbol{\nwarrow}$}}}}%
      \put(3840,6460){\rotatebox{-270}{\makebox(0,0){\strut{}{\large $\boldsymbol{\nwarrow}$}}}}%
      \put(4160,6380){\rotatebox{-270}{\makebox(0,0){\strut{}{\large $\boldsymbol{\nwarrow}$}}}}%
      \put(4520,6220){\rotatebox{-270}{\makebox(0,0){\strut{}{\large $\boldsymbol{\nwarrow}$}}}}%
    }%
    \gplgaddtomacro\gplfronttext{%
    }%
    \gplgaddtomacro\gplbacktext{%
      \colorrgb{0.50,0.50,0.50}%
      \put(385,1088){\makebox(0,0)[r]{\strut{}-2}}%
      \colorrgb{0.50,0.50,0.50}%
      \put(385,1511){\makebox(0,0)[r]{\strut{}-1.5}}%
      \colorrgb{0.50,0.50,0.50}%
      \put(385,1934){\makebox(0,0)[r]{\strut{}-1}}%
      \colorrgb{0.50,0.50,0.50}%
      \put(385,2358){\makebox(0,0)[r]{\strut{}-0.5}}%
      \colorrgb{0.50,0.50,0.50}%
      \put(385,2781){\makebox(0,0)[r]{\strut{} 0}}%
      \colorrgb{0.50,0.50,0.50}%
      \put(385,3204){\makebox(0,0)[r]{\strut{} 0.5}}%
      \colorrgb{0.50,0.50,0.50}%
      \put(385,3627){\makebox(0,0)[r]{\strut{} 1}}%
      \colorrgb{0.50,0.50,0.50}%
      \put(385,4051){\makebox(0,0)[r]{\strut{} 1.5}}%
      \colorrgb{0.50,0.50,0.50}%
      \put(385,4474){\makebox(0,0)[r]{\strut{} 2}}%
      \colorrgb{0.50,0.50,0.50}%
      \put(385,4897){\makebox(0,0)[r]{\strut{} 2.5}}%
      \colorrgb{0.50,0.50,0.50}%
      \put(556,732){\rotatebox{30}{\makebox(0,0)[l]{\strut{} 53500}}}%
      \colorrgb{0.50,0.50,0.50}%
      \put(1162,732){\rotatebox{30}{\makebox(0,0)[l]{\strut{} 54000}}}%
      \colorrgb{0.50,0.50,0.50}%
      \put(1767,732){\rotatebox{30}{\makebox(0,0)[l]{\strut{} 54500}}}%
      \colorrgb{0.50,0.50,0.50}%
      \put(2373,732){\rotatebox{30}{\makebox(0,0)[l]{\strut{} 55000}}}%
      \colorrgb{0.50,0.50,0.50}%
      \put(2979,732){\rotatebox{30}{\makebox(0,0)[l]{\strut{} 55500}}}%
      \colorrgb{0.50,0.50,0.50}%
      \put(3584,732){\rotatebox{30}{\makebox(0,0)[l]{\strut{} 56000}}}%
      \colorrgb{0.50,0.50,0.50}%
      \put(4190,732){\rotatebox{30}{\makebox(0,0)[l]{\strut{} 56500}}}%
      \csname LTb\endcsname%
      \put(0,2992){\rotatebox{-270}{\makebox(0,0){\strut{}$\Delta P_{\rm b} (\times 10^{-8}\ days)$}}}%
      \put(2548,500){\rotatebox{0}{\makebox(0,0){\strut{}MJD}}}%
      \put(1140,2620){\rotatebox{-270}{\makebox(0,0){\strut{}{\large $\boldsymbol{\swarrow}$}}}}%
      \put(1760,3020){\rotatebox{-270}{\makebox(0,0){\strut{}{\large $\boldsymbol{\nwarrow}$}}}}%
      \put(2180,3120){\rotatebox{-270}{\makebox(0,0){\strut{}{\large $\boldsymbol{\nwarrow}$}}}}%
      \put(2300,3420){\rotatebox{-270}{\makebox(0,0){\strut{}{\large $\boldsymbol{\nwarrow}$}}}}%
      \put(3840,3440){\rotatebox{-270}{\makebox(0,0){\strut{}{\large $\boldsymbol{\nwarrow}$}}}}%
      \put(4160,3070){\rotatebox{-270}{\makebox(0,0){\strut{}{\large $\boldsymbol{\nwarrow}$}}}}%
      \put(4520,3340){\rotatebox{-270}{\makebox(0,0){\strut{}{\large $\boldsymbol{\nwarrow}$}}}}%
    }%
    \gplgaddtomacro\gplfronttext{%
    }%
    \gplbacktext
    \put(0,0){\includegraphics{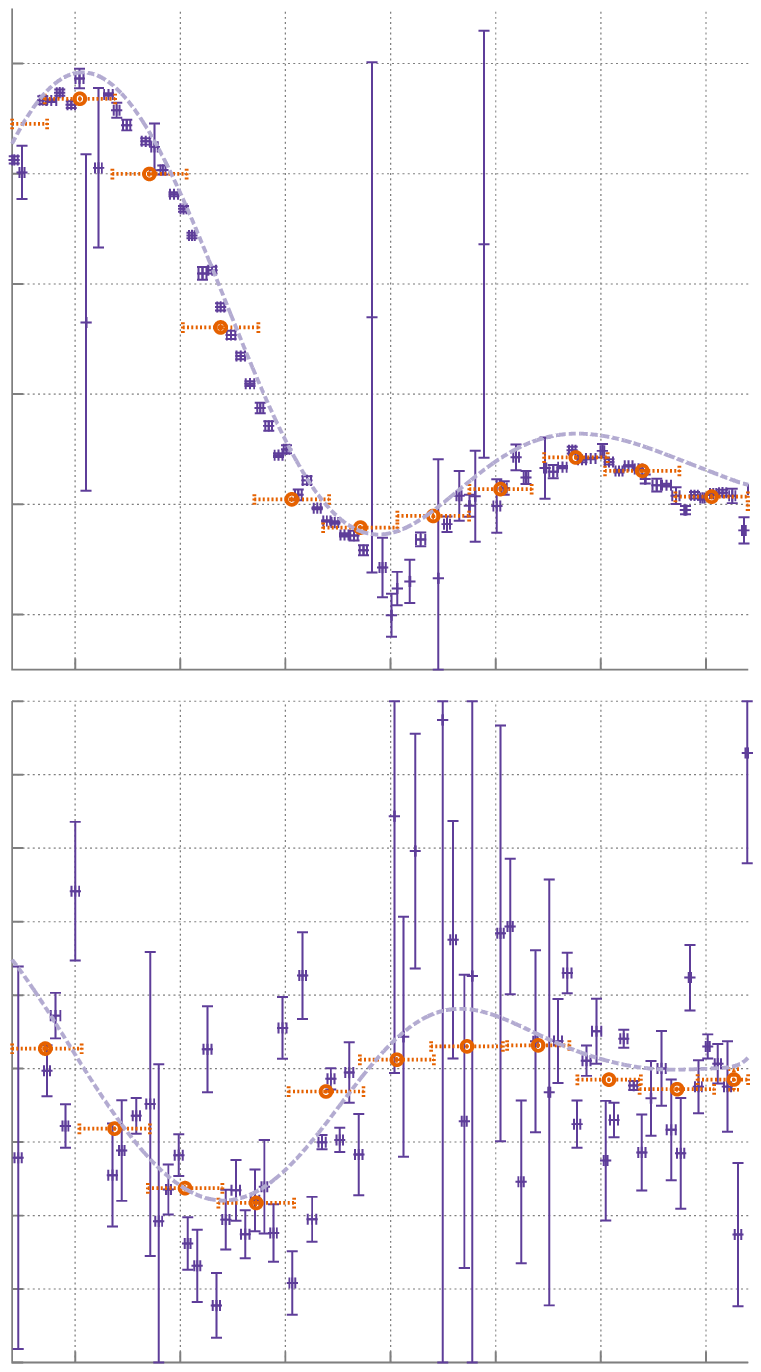}}%
    \gplfronttext
  \end{picture}%
\endgroup